\documentclass[12pt]{emulateapj}

\usepackage{amsmath,amssymb}  
\usepackage{graphicx}
\pdfoutput=1                  

\newcommand{\ngroup}{206 }      
\newcommand{\ogroup}{170 }      
\newcommand{\pgroup}{127 }      
\newcommand{\qgroup}{118 }      

\newcommand{\srmin}{40kpc }
\newcommand{\srmax}{4Mpc }

\newcommand{\lum}{h_{72}^{-2}\,\mathrm{erg}\,\mathrm{s}^{-1}}
\newcommand{\flux}{erg cm$^{-2}$ s$^{-1}$ }

\newcommand{\mlx}{$\rm{M} - \rm{L}_{\rm X}$ }
\newcommand{\lxm}{$\rm{L}_{\rm X} - \rm{M}$ }
\newcommand{\m}{$\rm{M}_{\rm 200}$ }
\newcommand{\lx}{$\rm{L}_{\rm X}$ }
\newcommand{\mass}{h_{72}^{-1} M_{\sun}}
 
 \newcommand{\eg}{{\it e.g.}}
 
 \newcommand{\ie}{{\it i.e.}}

 \def\multidrizzle{{\tt MultiDrizzle}}
 
\newcommand{\cosalpha}{$\alpha=0.66 \pm 0.14$}   
\newcommand{\henkalpha}{$\alpha=0.64 \pm 0.03$}
\newcommand{\sssig}{3.7$\sigma$}

\slugcomment{ }
 
\shorttitle{A Weak Lensing Study of X-ray Groups in COSMOS}
\shortauthors{A.\ Leauthaud}
 
\begin{document}
  

 \title{ A Weak Lensing Study of X-ray Groups in the COSMOS survey:
   Form and Evolution of the Mass-Luminosity Relation}



\author{Alexie Leauthaud\altaffilmark{1,2},
Alexis Finoguenov\altaffilmark{3,4},
Jean-Paul Kneib\altaffilmark{5},
James E. Taylor\altaffilmark{6}, 
Richard Massey\altaffilmark{7},
Jason Rhodes\altaffilmark{8,9},
Olivier Ilbert\altaffilmark{5},
Kevin Bundy\altaffilmark{10},
Jeremy Tinker\altaffilmark{2},
Matthew R. George\altaffilmark{11},
Peter Capak\altaffilmark{12},
Anton M. Koekemoer\altaffilmark{13},
David E. Johnston\altaffilmark{14},
Yu-Ying Zhang\altaffilmark{15},
Nico Cappelluti\altaffilmark{3},
Richard S. Ellis\altaffilmark{9},
Martin Elvis\altaffilmark{16},
Catherine Heymans\altaffilmark{7},
Oliver Le F\`{e}vre\altaffilmark{5},
Simon Lilly\altaffilmark{17},
Henry J. McCraken\altaffilmark{18},
Yannick Mellier\altaffilmark{18},
Alexandre R\'{e}fr\'{e}gier\altaffilmark{19},
Mara Salvato\altaffilmark{3,9},
Nick Scoville\altaffilmark{9},
George Smoot\altaffilmark{1,2},
Masayuki Tanaka\altaffilmark{20},
Ludovic Van Waerbeke\altaffilmark{21},
Melody Wolk\altaffilmark{22}}

\submitted{Accepted to ApJ}

\email{asleauthaud@lbl.gov}

\altaffiltext{1}{Lawrence Berkeley National Laboratory, 1 Cyclotron
  Road, Berkeley CA 94720}
\altaffiltext{2}{Berkeley Center for Cosmological Physics, University
  of California, Berkeley, CA 94720, USA}
\altaffiltext{3}{Max Planck Institut f\"{u}r extraterrestrische
  Physik, Giessenbachstrasse, D-85748 Garchingbei M\"{u}nchen,
  Germany}
\altaffiltext{4}{University of Maryland Baltimore County, 1000 Hilltop
  circle, Baltimore, MD 21250, USA}
\altaffiltext{5}{LAM, CNRS-UNiv Aix-Marseille, 38 rue F. Joliot-Curis,
  13013 Marseille, France}
\altaffiltext{6}{Department of Physics and Astronomy, University of
  Waterloo, 200 University Avenue West, Waterloo, Ontario, Canada N2L
  3G1}
\altaffiltext{7}{Institute for Astronomy, Blackford Hill, Edinburgh
  EH9 3HJ UK}
\altaffiltext{8}{Jet Propulsion Laboratory, California Institute of
  Technology, Pasadena, CA 91109}
\altaffiltext{9}{California Institute of Technology, MC 105-24, 1200
  East California Boulevard, Pasadena, CA 91125, USA}
\altaffiltext{10}{Hubble fellow, Department of Astronomy, University of
  California, Berkeley, CA 94720, USA}
\altaffiltext{11}{Department of Astronomy, University of
  California, Berkeley, CA 94720, USA}
\altaffiltext{12}{Spitzer Science Center, 314-6 Caltech, 1201
  E. California Blvd. Pasadena, CA, 91125, USA}
\altaffiltext{13}{Space Telescope Science Institute, 3700 San Martin
  Drive, Baltimore, MD 21218, USA}
\altaffiltext{14}{Department of Physics \& Astronomy, Northwestern
  University, Evanston, IL60208-2900, USA}
\altaffiltext{15}{Argelander-Institut f\"ur Astronomie, Universit\"at Bonn, Auf dem H\"ugel 71, 53121 Bonn, Germany}
\altaffiltext{16}{Harvard-Smithsonian Center for Astrophysics 60
  Garden St., Cambridge, Massachusetts 02138 USA}
\altaffiltext{17}{Institute of Astronomy, Department of Physics, ETH Zurich, CH-8093, Switzerland}
\altaffiltext{18}{Institut d'Astrophysique de Paris, UMR 7095, 98 bis
  Boulevard Arago, 75014 Paris, France}
\altaffiltext{19}{Service d'Astrophysique, CEA/Saclay, 91191
  Gif-sur-Yvette, France}
\altaffiltext{20}{European Southern Observatory,
  Karl-Schwarzschild-Str. 2, 85748 Garching bei Munchen, Germany}
\altaffiltext{21}{Department of Physics \& Astronomy, University of
  British Columbia, 6224 Agricultural Road, Vancouver, B.C. V6T 1Z1,
  Canada}
\altaffiltext{22}{Ecole normale sup\`{e}rieure de Cachan, 61 avenue du
  Pr\'{e}sident Wilson, 94235, Cachan cedex}

  
\begin{abstract}
  Measurements of X-ray scaling laws are critical for improving
  cosmological constraints derived with the halo mass function and for
  understanding the physical processes that govern the heating and
  cooling of the intracluster medium. In this paper, we use a sample
  of \ngroup X-ray selected galaxy groups to investigate the scaling
  relation between X-ray luminosity ($\rm{L}_{\rm X}$) and halo mass
  ($\rm{M}_{\rm 200}$) where $\rm{M}_{\rm 200}$ is derived via stacked
  weak gravitational lensing. This work draws upon a broad array of
  multi-wavelength COSMOS observations including 1.64 degrees$^2$ of
  contiguous imaging with the Advanced Camera for Surveys (ACS) to a
  limiting magnitude of $\rm{I}_{ \rm F814W}=26.5$ and deep {\sl
    XMM-Newton/Chandra} imaging to a limiting flux of $1.0\times
  10^{-15}$ \flux in the 0.5-2 keV band. The combined depth of these
  two data-sets allows us to probe the lensing signals of X-ray
  detected structures at both higher redshifts and lower masses than
  previously explored. Weak lensing profiles and halo masses are
  derived for nine sub-samples, narrowly binned in luminosity and
  redshift. The COSMOS data alone are well fit by a power law, ${\rm
    M}_{\rm 200} \propto (\rm{L}_{\rm X})^{\alpha}$, with a slope of
  \cosalpha . These results significantly extend the dynamic range for
  which the halo masses of X-ray selected structures have been
  measured with weak gravitational lensing. As a result, tight
  constraints are obtained for the slope of the \mlx relation. The
  combination of our group data with previously published cluster data
  demonstrates that the \mlx relation is well described by a single
  power law, \henkalpha, over two decades in mass, $\rm{M}_{\rm 200}
  \sim 10^{13.5}$ -- $10^{15.5} \mass$. These results are inconsistent
  at the \sssig\ level with the self-similar prediction of
  $\alpha=0.75$. We examine the redshift dependence of the \mlx
  relation and find little evidence for evolution beyond the rate
  predicted by self-similarity from $z \sim 0.25$ to $z \sim 0.8$.
\end{abstract}
 

 
 


\altaffiltext{$\star$}{Based on observations with the NASA/ESA {\em
    Hubble Space Telescope}, obtained at the Space Telescope Science
  Institute, which is operated by AURA Inc, under NASA contract NAS
  5-26555; also based on data collected at : the Subaru Telescope,
  which is operated by the National Astronomical Observatory of Japan;
  the XMM-Newton, an ESA science mission with instruments and
  contributions directly funded by ESA Member States and NASA; the
  European Southern Observatory under Large Program 175.A-0839, Chile;
  Kitt Peak National Observatory, Cerro Tololo Inter-American
  Observatory, and the National Optical Astronomy Observatory, which
  are operated by the Association of Universities for Research in
  Astronomy, Inc. (AURA) under cooperative agreement with the National
  Science Foundation; the National Radio Astronomy Observatory which
  is a facility of the National Science Foundation operated under
  cooperative agreement by Associated Universities, Inc ; and and the
  Canada-France-Hawaii Telescope with MegaPrime/MegaCam operated as a
  joint project by the CFHT Corporation, CEA/DAPNIA, the National
  Research Council of Canada, the Canadian Astronomy Data Centre, the
  Centre National de la Recherche Scientifique de France, TERAPIX and
  the University of Hawaii.}


\section{Introduction}
Groups and clusters of galaxies, formed through the gravitational
collapse of massive dark matter halos, are now readily identified up
to redshift one and even beyond
\citep[\eg][]{Stanford:2006,Eisenhardt:2008}. Baryonic tracers such as
red-sequence galaxies, typically abundant at the centers of groups and
clusters, or X-ray emission from the hot intracluster medium (ICM),
have proved to be especially successful in this task
\citep[\eg][]{Gladders:2005,Koester:2007,Finoguenov:2007}. Nonetheless,
these observables only trace the tip of the iceberg given that the
vast majority of the underlying mass is in the form of dark matter.

The quantification of the total mass (both dark and baryonic) of
groups and clusters of galaxies is an important endeavour from both a
cosmological and an astronomical standpoint. In particular, several
lines of research would benefit from a clearer understanding of the
relationship between the total halo mass of groups and clusters and
their baryonic tracers. We outline several briefly here \citep[for a
recent review on this subject see][]{Voit:2005}. From a cosmological
perspective, the number density of groups and clusters as a function
of total mass is of fundamental interest because it is sensitive to
both the expansion and growth history of the universe and can be used
to constrain cosmological parameters such as $\Omega_m, \sigma_8,$ and
$\Omega_{\Lambda}$
\citep[\eg][]{White:1993,Wang:1998a,Haiman:2001,Rozo:2004, Wang:2004a,
  Bahcall:2004, Rozo:2009}. Furthermore, modifications to the laws of
gravity which can be invoked as a possible physical mechanism for
acceleration, could imprint telltale signatures in the abundance and
dark matter structure of groups and clusters of galaxies
\citep[][]{Rapetti:2008,Schmidt:2008}. Unfortunately, the common
difficulty encountered with each of these enquires is that theories
and simulations make dark matter based predictions but our most
accessible observables (such as richness or X-ray luminosity) are
baryonic in nature.

It has long been recognized that baryonic observables are subject to
complex and poorly understood physical processes that make them
imperfect dark matter tracers. For example, X-ray studies discovered
early on that the theory of pure gravitational collapse which makes
simple predictions for the shape and amplitude of X-ray scaling
relations \citep[also known as the {\it self-similar
  model,}][]{Kaiser:1986}, fails to match observations such as the
slope and normalization of the \lx-T relation \citep[][and references
therein]{Voit:2005} implying that other non-gravitational (and still
much debated) processes have significantly affected the thermodynamic
state of the ICM. Additional heating and cooling mechanisms that are
invoked to solve this puzzle lead to different predictions for the
shape and redshift evolution of X-ray scaling relations. On the one
hand, the fact that X-ray scaling relations deviate from simple models
is a plague for cosmologists because there is no straightforward
recipe to estimate total halo masses. On the other hand, from an
astronomical perspective, the comparison between X-ray observables and
total halo mass contains valuable clues about the physical processes
that govern galaxy formation and the heating and cooling of the ICM.

For all of these reasons, more precise measurements of the mean and
scatter in the relationship between total halo mass and various
baryonic tracers of groups and clusters of galaxies are highly
desirable (\eg\ see discussions in \citet{Voit:2005} and
\citet{Albrecht:2006}).

At present, there are five popular methods for detecting groups and
clusters of galaxies: a) optical detection via the red-sequence
\citep[\eg][]{Gladders:2005,Koester:2007}, b) detection via the
Sunyaev-Zeldovich (SZ) effect which measures the distortion of the CMB
spectrum due to the hot ICM
\citep[\eg][]{Sunyaev:1970,Sunyaev:1972,Carlstrom:2002,Benson:2004,Staniszewski:2008},
c) detection via X-ray emission
\citep[\eg][]{Bohringer:2000,Hasinger:2001,Finoguenov:2007,
  Vikhlinin:2008}, d) spectroscopic identification
\citep[\eg][]{Gerke:2005,Miller:2005,Knobel:2009}, and e) detection
via weak lensing maps
\citep[\eg][]{Marian:2006,Miyazaki:2007,Massey:2007a}. This last
technique is the simplest in terms of the underlying physics and is
the only method for which the total halo mass can be directly probed,
independently of both the baryons and the dynamical state of the
cluster. However, shear maps can only detect the most massive systems
($M>10^{14} M_{\odot}$) and are limited to moderate redshifts because
the lensing weight function peaks mid-way between the source and the
observer, with galaxy shapes increasingly difficult to measure at
$z>1$. X-ray observations on the other hand, can more simply probe
complete samples of groups and clusters, but departures from virial
equilibrium and non-thermal pressure components in the ICM can bias
X-ray based hydrostatic mass estimates \citep[\eg][]{Nagai:2007}. The
SZ effect has the attractive property of being redshift independent,
and the integrated SZ flux increment, Y, may be less sensitive to the
baryon physics of cluster cores \citep[][]{Motl:2005,Nagai:2006} but
mass measurements with the SZ effect face other challenges such as the
identification and removal of radio point-sources
\citep[][]{Vale:2006}, sky confusion owing to projection effects
\citep{White:2002a}, and possibly a larger scatter in the Y-M relation
than previously estimated due to feedback processes
\citep[][]{Shaw:2008}.

Given these considerations, a promising strategy is to employ a robust
and efficient cluster detection method \citep[to which the ultimate
solution may be a combination of several techniques such as described
in][]{Cohn:2009} and to perform an absolute mass calibration of
baryonic tracers via weak gravitational lensing
\citep[][]{Hoekstra:2007a,Rykoff:2008,Berge:2008}.

The focus of this paper is to advance these goals by calibrating the
slope and amplitude of the \mlx relation for galaxy groups using
cross-correlation weak lensing in the COSMOS survey (also called
``group-galaxy lensing''). Extending weak lensing measurements into
the group regime is particularly important in order to extend the
dynamic range of weak-lensing-based mass-estimates so as to more
accurately determine the slopes of scaling relations. Using the COSMOS
sample, we show that X-ray detections span a more complete and wide
range of redshift and mass than detections via shear maps. Indeed,
high redshift and small structures are challenging to detect directly
with shear because of the shape of the lensing weight function (see
$\S$\ref{group_selection}). Nevertheless, once they have been
identified, groups and clusters can be studied via stacking
techniques. A notable advantage of this method is that measurements
are unaffected by uncorrelated mass along the line-of-sight whereas
mass estimates for individually detected clusters are subject to $\sim
20 \%$ uncertainties
\citep[][]{Metzler:2001,Hoekstra:2003a,de-Putter:2005}. The associated
drawback with stacking is that the intrinsic scatter around the mean
relation is difficult to recover.

In order to employ the stacked weak lensing technique, tight baryonic
tracers of halo mass are highly desirable. The X-ray luminosity of
groups and clusters is considered to be a reasonable tracer of halo
mass with a logarithmic scatter in the \mlx relation of roughly $20\%$
to $30\%$ \citep[][]{Stanek:2006, Maughan:2007, Pratt:2008, Rozo:2008,
  Rykoff:2008, Vikhlinin:2009}. A large fraction of this scatter has
been shown to be associated with the presence of cool-cores and simple
excision techniques can reduce the scatter to sub-$20\%$ levels
\citep[][]{Maughan:2007,Pratt:2008}. Although more tightly correlated
mass tracers have been identified such as the ${\rm Y_{X}}$ parameter
\citep[\eg][]{Kravtsov:2006} -- such indicators require the
measurement of an X-ray spectrum which is not possible for most survey
data where count rates are low. Our choice of \lx as a mass proxy
reflects that fact that it is a simple X-ray observable, accessible
with survey quality data, and the only one that can be easily measured
at high redshift. Temperature measurements may be feasible for a small
fraction of high redshift objects but cosmological studies that
require complete samples of high redshift groups and clusters will
need simple mass proxies like ${\rm{L_{\rm X}}}$. The details of the
\mlx relation are also important (regardless of the choice of a mass
proxy) for determining effective volumes as a function of mass in
X-ray flux limited surveys \citep[][]{Stanek:2006, Vikhlinin:2009}.

The layout of this paper is as follows. The data are presented in
$\S$\ref{cosmos_survey} and the theoretical lensing background is
developed in $\S$\ref{lensing_theory}. The construction of the group
catalog and the lens selection are specified in
$\S$\ref{group_selection}. Details regarding the adopted form of the
\mlx relation are given in $\S$\ref{lx_m_relation}. Our main results
are presented in $\S$\ref{results} followed by our assessment of the
systematic errors in $\S$\ref{syst_error}. Finally, we discuss the
results and draw up our conclusions in $\S$\ref{discussion}.

We assume a WMAP5 $\Lambda$CDM cosmology with $\Omega_{\rm m}=0.258$,
$\Omega_\Lambda=0.742$, $\Omega_{\rm b}h^2=0.02273$, $n_{\rm
  s}=0.963$, $\sigma_{8}=0.796$, $H_0=72$ $h_{72}$
km~s$^{-1}$~Mpc$^{-1}$ \citep[][]{Hinshaw:2009}. All distances are
expressed in physical units of $h_{72}^{-1}$ Mpc. X-ray luminosities
are expressed in the 0.1-2.4 keV band, rest-frame. The letter M
denotes halo mass in general whereas \m is explicitly defined as
$M_{200}\equiv M(<r_{200})=200\rho_{crit}(z) \frac{4}{3}\pi r_{200}^3$
where $r_{200}$ is the radius at which the mean interior density is
equal to 200 times the critical density ($\rho_{crit}\equiv
3H^2(z)/8\pi G$). The function $E(z) \equiv H(z)/H_0 =
\sqrt{\Omega_{m}(1+z)^3+\Omega_{\Lambda}}$ represents the Hubble
parameter evolution for a flat metric. All magnitudes are given on the
AB system.


\section{The COSMOS Survey}\label{cosmos_survey}

The COSMOS survey brings together a broad array of panchromatic
observations with imaging data from X-ray to radio wavelengths and a
large spectroscopic follow-up program (zCOSMOS) with the VLT
\citep{Scoville:2007,Koekemoer:2007,Lilly:2007}. In particular, the
COSMOS program has imaged the largest contiguous area (1.64
degrees$^2$) to date with the Hubble Space Telescope (HST) using the
Advanced Camera for Surveys (ACS) Wide Field Channel (WFC). In
addition to the ACS/WFC ($\rm{I}_{F814W}$) imaging, the COSMOS field
has been targeted by both the {\sl XMM-Newton} \citep[1.5
Ms,][]{Hasinger:2007,Cappelluti:2009} and the {\sl Chandra}
observatories \citep[1.8 Ms,][]{Elvis:2009}. The combination of ACS
imaging to provide accurate shape measurements, and of {\sl
  XMM-Newton/Chandra} imaging, sets the stage for the study of the
dark matter halos of galaxy groups via weak lensing techniques. In the
following sections, we describe the various data-sets and catalogs
employed in this analysis.

\subsection{ACS Lensing data}

Our general scheme for the construction of the COSMOS ACS lensing
catalog is based on \citet{Leauthaud:2007} and we refer the reader to
this publication for details regarding the source extraction and
catalog construction -- only a brief review will be given here. Since
\citet{Leauthaud:2007}, however, we have made a key improvement
regarding the effects of Charge Transfer Inefficiency (CTI) in the ACS
CCDs. This aspect in particular is outlined below in greater detail.

In \citet{Leauthaud:2007} and \citet{Rhodes:2007}, we remarked that
the COSMOS ACS images are strongly affected by CTI. As charge is
transferred during the CCD read-out process, a certain fraction is
retained by charge traps (created by cosmic ray hits) in the
pixels. This causes flux to be trailed behind objects as the traps
gradually release their charge, spuriously elongating them in a
coherent direction that mimics a lensing signal. Our previous work
employed a parametric scheme to correct for this effect at a catalog
level. Although a parametric scheme provides a first order correction
of CTI, it neglects the dependence of the CTI on object size, radial
profile, and shape for example. For this reason, in
\citet[][]{Massey:2009}, we have developed a physically motivated
correction scheme that operates on the raw data and that has been
shown to achieve a 97$\%$ level of correction. Using this scheme, we
have produced a new set of raw ``unrotated'' ACS/WFC data (version
2.0) in which the CTI trailing is reduced by more than an order of
magnitude \citep[for further details, see][]{Massey:2009}. The raw
data are co-added using the same \multidrizzle\ set-up as in our
previous work.

We use Version 2.5.0 of the SExtractor photometry package
\citep{Bertin:1996} to extract a source catalog of positions from the
v2.0 ACS data using the same \textit{``Hot-Cold''} method as in
\citet{Leauthaud:2007}. Defects and diffraction spikes are carefully
removed from the catalog, leaving a total of $ \sim 1.2 \times 10^6$
objects to a limiting magnitude of $\rm{I}_{F814W}=26.5$.

The next step is to measure the shapes of galaxies and to correct them
for the distortion induced by the time varying ACS PSF as described in
\citet[][]{Rhodes:2007}. As compared to \citet[][]{Rhodes:2007}, the
parametric CTI correction is no longer applied because this effect has
already been removed in the raw images.

Simulated images are used to derive the shear susceptibility factors
that are necessary in order to transform shape measurements into
unbiased shear estimators. Finally, for every galaxy we derive a shape
measurement error and utilize this quantity to extract the intrinsic
shape noise of the galaxy sample.  Representing a number density of
$66$ galaxies/arcminute$^{2}$, the final COSMOS weak lensing catalog
contains $3.9 \times 10^5$ galaxies with accurate shape measurements.

\subsection{Photometric and Spectroscopic Redshifts}\label{photoz}

Redshift information is critical for both the lens and source
populations because it allows one to correctly scale lensing
relations, to remove foreground contamination, and to study weak
lensing signals in terms of physical instead of angular distances. We
use an updated and improved version of the photometric redshifts
(hereafter photozs) presented in \citet{Ilbert:2009} which have been
computed with over 30 bands of multi-wavelength data. The main
differences between the \citet{Ilbert:2009} catalog and the one that
we use here is the addition of deep $H$ band data and small
improvements in the template fitting technique. Details regarding the
data and the photometry can be found in \citet{Capak:2007}.

Photozs were estimated using a $\chi^2$ template fitting method (Le
Phare) and compared with large spectroscopic samples from the Very
Large Telescope (VLT) Visible Multi-Object Spectrograph (VIMOS)
\citep[][]{Lilly:2007} and the Keck Deep Extragalactic Imaging
Multi-Object Spectrograph (DEIMOS). The combined spectroscopic
redshift sample comprises: 10801 galaxies at $z\sim0.48$, 696 at
$z\sim 0.74$, and 870 at $z\sim 2.2$. The dispersion in the photozs as
measured by comparing to the spectroscopic sample is $\sigma_{\Delta
  z/(1+{\rm zspec})}=0.007$ at $i^{+}_{AB}<22.5$ where $\Delta z =
z_{\rm spec} - z_{\rm phot}$.  The deep IR and Infrared Array Camera
(IRAC) data enables the photozs to be calculated even at fainter
magnitudes with a reasonable accuracy of $\sigma_{\Delta z/(1+{\rm
    zspec})}=0.06$ at $i^{+}_{AB} \sim 24$. In particular, deep $J$,
$H$, $K$, and $u^{*}$ band data allow for a better estimation of the
photozs at $z>1$ via the 4000\AA\ break which is shifted into the
infrared (IR).

Larger samples of spectroscopic redshifts at $z>1$ will ultimately be
required to define the most trustworthy magnitude and redshift range
for the source galaxies (in a similar fashion to
\citet{Mandelbaum:2008}). Meanwhile, to mitigate the effects of photoz
uncertainties, we use a conservative source galaxy selection which
will be discussed in more detail in $\S$\ref{syst_error}. In short, we
reject all source galaxies with a secondary peak in the redshift
probability distribution function (\ie\ the parameter {\sc zp\_sec} is
non zero in the \citet{Ilbert:2009} catalog). This cut is aimed to
reduce the number of catastrophic errors (a preferential shift in a
certain population of galaxies from one redshift bin to another) in
the source catalog. The {\sc zp\_sec$>0$} galaxy population is
expected to contain a large fraction of catastrophic errors
\citep[roughly 40\%-50\%,][]{Ilbert:2006,Ilbert:2009}. The photoz
quality cuts reduce the number density of source galaxies from $66$ to
$34$ galaxies/arcminute$^{2}$. The final mean redshift and magnitude
of the source sample is $\langle z \rangle \sim 1$ and $\langle
\rm{I}_{F814W} \rangle \sim 24$. In addition to these quality cuts,
for each lens-source pair, we demand that $z_{\rm source}-z_{\rm
  lens}> \sigma_{68\%}(z_{\rm source})$ and that $z_{\rm
  source}-z_{\rm lens}> 0.1$ to ensure a clean selection of background
galaxies.

\subsection{Stellar Mass Estimates}\label{stellar_masses}

Stellar masses are used to identify the Most Massive Central Galaxy
(MMCG) (see $\S$\ref{center_determination} for more details) and are
estimated using the Bayesian code described in Bundy (2006a) and Bundy
et al. (2006b). Briefly, an observed galaxy's spectral energy
distribution (SED) and photoz is referenced to a grid of models
constructed using the \citet{bruzual:2003} synthesis code.  The grid
includes models that vary in age, star formation history, dust
content, and metallicity.  At each grid point, the probability that
the observed SED fits the model is calculated, and the corresponding
stellar-mass to K-band luminosity ratio and stellar mass is stored.
By marginalizing over all parameters in the grid, the stellar mass
probability distribution is obtained.  The median of this distribution
is taken as the stellar mass estimate, and the width encodes the
uncertainty due to degeneracies and uncertainties in the model
parameter space.  The final uncertainty on the stellar mass also
includes the K-band photometry uncertainty as well as the expected
error on the luminosity distance that results from the photoz
uncertainty.  The typical final uncertainty is 0.2-0.3 dex.

\subsection{XMM and Chandra data}

The entire COSMOS region has been mapped through 54 overlapping {\sl
  XMM-Newton} pointings and additional {\sl Chandra} observations
cover the central region (0.9 degrees$^2$) to higher resolution. A
composite {\sl XMM-Newton} and {\sl Chandra} mosaic has been used to
detect and measure the fluxes of groups and clusters to a $4\sigma$
detection limit\footnote[1]{Quoted detection limits correspond to the
  wavelet scale-wise reconstruction. See \citet{Finoguenov:2007} for
  more details.} of $1.0\times 10^{-15}$ \flux over $96\%$ of the ACS
field. The general data reduction process can be found in
\citet{Finoguenov:2007} and details regarding improvements and
modifications to the original catalog are given in
$\S$\ref{group_selection}. In particular, the group and cluster
catalog used in this paper features a more conservative point-source
removal procedure than in \citet{Finoguenov:2007}. Redshift
identification has also improved thanks to the increased photoz
accuracy and to the availability of more spectroscopic data (see
$\S$\ref{redshift_determination}). In total, the catalog used in this
paper contains \ngroup X-ray groups and clusters of galaxies over 1.64
degrees$^2$, spanning the redshift range $0<z<1.6$ and with a
rest-frame 0.1--2.4 keV luminosity range between $10^{41}$ and
$10^{44}$ erg s$^{-1}$.


\section{Theoretical Lensing Framework}\label{lensing_theory}

\subsection{From galaxy shapes to $\Delta\Sigma$}

In the weak gravitational lensing limit, the observed shape
$\varepsilon_{\rm obs}$ of a source galaxy is directly related to the
lensing induced shear $\gamma$ according to

\begin{equation}
  \varepsilon_{\rm obs} = \varepsilon_{\rm int}+\gamma,
\label{ellipticity_shear}
\end{equation}

\noindent where $\varepsilon_{\rm int}$ is the source galaxy's
intrinsic shape that would be observed in the absence of gravitational
lensing. In our notation, $\varepsilon_{\rm int}$, $\varepsilon_{\rm
  obs}$, and $\gamma$ are spin-2 tensors. The above relationship
indicates that galaxies would be ideal tracers of the distortions
caused by gravitational lensing if the intrinsic shape
$\varepsilon_{\rm int}$ of each source galaxy was known \textit{a
  priori}. However, lensing measurements exhibit an intrinsic
limitation, encoded in the width of the ellipticity distribution of
the galaxy population, noted here as $\sigma_{\rm int}$, and often
referred to as the ``intrinsic shape noise''. Because the intrinsic
shape noise \citep[of order $\sigma_{\rm int} \sim
0.27$,][]{Leauthaud:2007} is significantly larger than $\gamma$
(typically $\gamma \sim 0.05$ for this work), shears must be estimated
by averaging over a large number of source galaxies.

Throughout this paper, the gravitational shear is noted as $\gamma$
whereas $\tilde\gamma$ represents our estimator of $\gamma$. The
uncertainty in the shear estimator is a combination of unavoidable
intrinsic shape noise, $\sigma_{\rm int}=\sqrt{\langle
  \varepsilon_{\rm int}^2\rangle}$, and of shape measurement error,
$\sigma_{\rm meas}$:

\begin{equation}\label{eq:3}
  \sigma_{\rm \tilde\gamma}^2 = \sigma_{\rm int}^2+\sigma_{\rm meas}^2.
\end{equation}
 
We will refer to $\sigma_{\rm \tilde\gamma}$ as ``shape noise''
whereas $\sigma_{\rm int}$ will be called the ``intrinsic shape
noise''. The former includes the shape measurement error and will vary
according to each specific data-set and shape measurement
method. Averaged over the whole COSMOS field, the weak lensing
distortions represent a negligible perturbation to Equation
\ref{eq:3}. The intrinsic shape noise and measurement error for COSMOS
have been characterized in \citet[][]{Leauthaud:2007} by using a
sample of 27000 source galaxies that lie within the overlapping
regions of adjacent pointings. The shape measurement error is
determined for every source galaxy as a function of size and
magnitude. For this paper, an intrinsic shape noise of $\sigma_{\rm
  int}=0.27$ is assumed.

The derivation of our shear estimator is presented in
\citet[][]{Leauthaud:2007}. We employ the RRG method \citep[see][for
further details]{Rhodes:2000} for galaxy shape measurements. Briefly,
we form $\tilde\gamma$ from the PSF corrected ellipticity according to

\begin{equation}
  \tilde\gamma=C \times \frac{\varepsilon_{\rm obs}}{G},
\end{equation}

\noindent where the shear susceptibility factor\footnote[2]{Not to be
  confused with Newton's constant which we have noted $G_{{ {\rm
        N}}}$.}, $G$, is measured from moments of the global
distribution of $\varepsilon_{\rm obs}$ and other, higher order shape
parameters \citep[see equation 28 in][]{Rhodes:2000}. Using a set of
simulated images similar to those of Shear TEsting Program
\citep[STEP;][]{Heymans:2006a,Massey:2006} but tailored exclusively to
this data-set, we find that, in order to correctly measure the input
shear on COSMOS-like images, the RRG method requires an overall
calibration factor of $C=(0.86^{+0.07}_{-0.05})^{-1}$ \citep[see][for
more details]{Leauthaud:2007}.

The shear signal induced by a given foreground mass distribution on a
background source galaxy will depend on the transverse proper distance
between the lens and the source and on the redshift configuration of
the lens-source system. A lens with a projected surface mass density,
$\Sigma(r)$, will create a shear that is proportional to the {\em
  surface mass density contrast}, $\Delta\Sigma(r)$:

\begin{equation}
  \Delta \Sigma(r)\equiv\overline{\Sigma}(< r)-\overline{\Sigma}(r)=\Sigma_{\rm crit}\times\gamma_t(r).
\label{dsigma}
\end{equation}

Here, $\overline{\Sigma}(< r)$ is the mean surface density within
proper radius $r$, $\overline{\Sigma}(r)$ is the azimuthally averaged
surface density at radius $r$
\citep[\eg][]{Miralda-Escude:1991,Wilson:2001}, and $\gamma_t$ is the
tangentially projected shear. The geometry of the lens-source system
intervenes through the {\em critical surface mass density},
$\Sigma_\mathrm{crit}$, which depends on the angular diameter
distances to the lens ($D_{\rm OL}$), to the source ($D_{\rm OS}$),
and between the lens and source ($D_{\rm LS}$):

\begin{equation}
  \Sigma_\mathrm{crit} = \frac{c^2}{4\pi G_{{{\rm N}}}}\,
  \frac{D_\mathrm{OS}}{D_\mathrm{OL}\,D_\mathrm{LS}}\;,
\label{sigma_crit}
\end{equation}

\noindent where $G_{\rm N}$ represents Newton's constant. Hence, if
  redshift information is available for every lens-source pair, each
  estimate of $\gamma_{t}$ can be directly converted to an estimate of
  $\Delta\Sigma$ which is a more desirable quantity because it
  depends only on the mass distribution of the lens.

To measure $\Delta\Sigma(r)$ with high signal-to-noise, the lensing
signal must be stacked over many foreground lenses and background
sources. For every $i$th lens and $j$th source separated by a proper
distance $r_{ij}$, an estimator of the mean excess projected surface
mass density at $r_{ij}$ is computed according to:

\begin{equation}
  \Delta \tilde\Sigma_{ij}(r_{ij})=\tilde\gamma_{t,ij}\times \Sigma_{{\rm crit},ij},
\label{dsigma5}
\end{equation}

\noindent where $\tilde\gamma_{t,ij}$ is the tangential shear of the
source relative to the lens. The COSMOS photometric redshifts
described in $\S$\ref{photoz} are used to estimate $\Sigma_{{\rm
    crit},ij}$ for every lens-source pair. In order to optimize the
signal-to-noise, an inverse variance weighting scheme is employed when
$\Delta\Sigma_{ij}$ is summed over many lens-source pairs. Each
lens-source pair is attributed a weight that is equal to the estimated
variance of the measurement:

\begin{equation}
  w_{ij} = \frac{1}{ \left( \Sigma_{{\rm crit},ij} \times \sigma_{\tilde\gamma,ij} \right)^2}. 
\end{equation}

In this manner, faint small galaxies which have large measurement
errors are down-weighted with respect to sources that have well
measured shapes.

In general, for the types of lenses studied in this paper (groups and
low mass clusters of galaxies), the signal-to-noise per lens is not
high enough to measure $\Delta\Sigma$ on an object by object basis so
instead we stack the signal over many lenses. For a given sample of
lenses, the total excess projected surface mass density is the
weighted sum over all lens-source pairs:

\begin{equation}
  \Delta\Sigma =  
  {\sum_{j=1}^{N_{Lens}} \sum_{i=1}^{N_{Source}} w_{ij} \times \tilde\gamma_{t,ij}\times \Sigma_{{\rm crit},ij}
    \over \sum_{j=1}^{N_{Lens}} \sum_{i=1}^{N_{Source}}w_{ij}} ~.
\end{equation}

\subsection{Non-Weak Shear}\label{non_weak_shear}

Equation \ref{ellipticity_shear} and subsequent derivations only hold
in the weak gravitational lensing limit, that is to say when $\gamma
\ll 1$ and $\kappa \ll 1$ ($\kappa = \Sigma / \Sigma_{\rm crit}$ is
the {\em convergence}). In reality, galaxy shapes trace the reduced
shear, $g=\gamma/(1-\kappa)$. The masses of the groups that we are
probing are such that the weak lensing assumption can begin to break
down in the most inner bins ($r<100$kpc h$_{72}^{-1}$) and for high
redshift source galaxies. In this regime, $g \sim \gamma$ is no longer
a valid assumption. Following the methodology outlined in
\citet[][]{Johnston:2007} and \citet[][]{Mandelbaum:2006}, it can be
shown that our weighted estimator for $\Delta\Sigma$ will have a
second-order contribution:

\begin{equation}
\Delta \tilde\Sigma = \Delta\Sigma + \Delta\Sigma~\Sigma~L_{\rm Z},
\end{equation}

\begin{equation}
  L_{\rm Z} = \frac{\langle \Sigma_{\rm crit}^{-3}\rangle }{\langle \Sigma_{\rm crit}^{-1}\rangle}.
\end{equation}

For further details, see Equation 19 in \citet[][]{Johnston:2007} and
Appendix A in \citet[][]{Mandelbaum:2006}. In a similar fashion to
\citet[][]{Johnston:2007}, we ignore the variations of $L_{\rm Z}$
within various radial bins. However, because our lens sample spans a
large redshift range, we do not use a constant value of $L_{\rm Z}$
for all redshift bins. Instead, $L_{\rm Z}$ is calculated from the
data for each redshift bin. Our values for $L_{\rm Z}$ are given in
Table 1.

\subsection{A Model for Predicting $\Delta\Sigma$}\label{model_for_ds}

A halo model approach is used to model the surface mass density
contrast $\Delta\Sigma$ as a function of transverse separation
\citep[\eg][]{Mandelbaum:2005a,Mandelbaum:2006,Mandelbaum:2006c,Yoo:2006,Johnston:2007}. The
total signal is modelled as the sum of two distinct components that
dominate the signal at different scales. The first term is due to the
baryonic mass contained within the central galaxy and only contributes
at scales below $\sim 50$kpc. The second term dominates the signal on
intermediate to large scales ($\sim 50$kpc to a few Mpc) and
represents the group-scale dark matter halo (also known as the
``one-halo term''). On the largest scales (above several Mpc), the
clustering of halos among themselves produces a contribution to
$\Delta\Sigma$ via the so-called ``two-halo term''. However, we have
found that the two-halo term is mostly sub-dominant at the scales that
we probe. We have tested that the exclusion of the two-halo term has
no impact on the results of this study and we therefore neglect this
term hereafter.

The baryonic mass of the central galaxies can have a non negligible
contribution to $\Delta\Sigma$ at small transverse separations from
the center of the stacked ensemble. Although the baryons typically
follow a Sersic profile, at the scales of interest for this study,
well above a few effective radii ($>$40kpc), the lensing
contribution of the baryons can be modeled by a simple point-source,
scaled to $\langle M_{CG}\rangle$, the average stellar mass of the
central galaxies (also see $\S$\ref{stellar_masses} and
$\S$\ref{center_determination}):

\begin{equation}
  \Delta\Sigma_{\rm stellar}(r)=\frac{\langle M_{CG}\rangle }{\pi r^2}.
\end{equation}

To be more precise, the baryons that have not yet transformed into
stars should also be considered. However, the majority of non-stellar
group baryons are in the form of diffuse hot gas spread throughout the
halo, in rough equilibrium with the dark matter potential. To first
order, the gas mass contribution should follow the dark matter
distribution.

We assume that the density profiles of dark matter halos follow {\sc
  nfw} profiles \citep[][]{Navarro:1997}. In this work, halo mass is
defined as $M_{200}\equiv M(<r_{200})=200\rho_{crit}(z) \frac{4}{3}\pi
r_{200}^3$ and $C_{200}$ denotes halo concentration. Numerous studies
have demonstrated that the mass, concentration, and characteristic
formation epoch of dark matter halos are closely linked and on
average, smaller halos tend to have higher concentrations
\citep[][]{Bullock:2001,Wechsler:2002,Maccio:2007,Zhao:2008}. For this
study, we adopt the \citet{Zhao:2008}\footnote[3]{see
  http://www.shao.ac.cn/dhzhao/mandc.html} mass-concentration relation
for a WMAP5 cosmology. By adopting this relation, $\Delta\Sigma_{\rm
  nfw}$ is fully described by two parameters, namely $M_{200}$ and
redshift.  Analytical formulas for the $\Delta\Sigma$ corresponding to
a {\sc nfw} profile can be found in \citet{Wright:2000} and in
\citet[][]{Bartelmann:1996}.

The additional gravitational potential due to the baryons is expected
to modify the density profiles of dark matter halos via adiabatic
contraction (Gnedin 2004, 2005, Sellwood and McGaugh
2005). Nevertheless, \citet{Mandelbaum:2006} have shown that this has
a negligible effect on the lensing signal on the scales that we
consider (above 40kpc) and we neglect this effect for this work.

The final model that we use for the weighted estimate $\Delta \tilde
\Sigma$ is:

\begin{equation}\label{ds_model}
\Delta \tilde \Sigma = \Delta\Sigma_{\rm stellar}+\Delta\Sigma_{\rm nfw}+\Delta\Sigma_{\rm nfw} \Sigma_{\rm nfw} L_{\rm Z}.
\end{equation}

We have included the second order contribution to $\Delta \tilde
\Sigma$ from non-weak shear. Note that only the dark matter halo
contributes to this term because $\Sigma_{\rm stellar}$ is zero at the
scales that we probe.


\section{X-ray Group Selection}\label{group_selection}

\subsection{X-ray Selection Versus Shear Maps}

\begin{figure*}[htb]
\centerline{\includegraphics[scale=0.6]{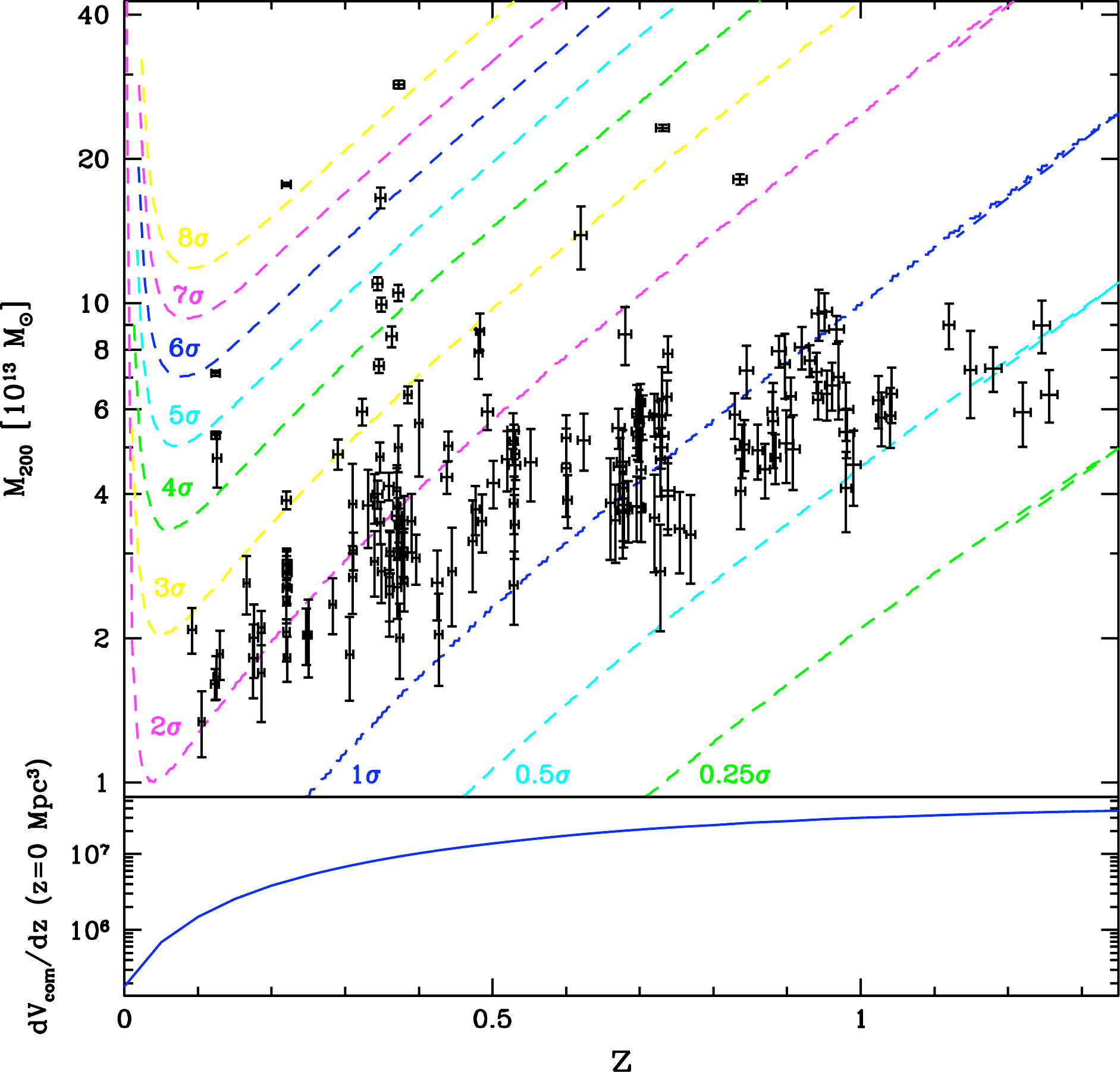}}
\caption{Upper panel: theoretically computed lensing detection
  significance (dashed curves) of X-ray structures (data-points) in
  the COSMOS field as a function of mass and redshift. Lensing
  detection significance values have been derived with the method
  outlined in \protect\citet{Hamana:2004} assuming a smoothed COSMOS
  redshift distribution and a source density of 66
  galaxies/arcminute$^2$. The predictions in this Figure represent a
  realistic upper limit for all near-future weak-lensing surveys,
  including space-based missions such as JDEM and EUCLID. Indeed, the
  survey depth of these missions is unlikely to exceed the current
  depth of COSMOS. Direct structure detection via shear maps is
  limited by the lensing weight function at high redshifts and at low
  masses. To identify such structures, one must resort to other
  detection techniques. Lower panel: comoving volume probed by the
  (1.64 degrees$^2$) survey per unit redshift. Although the low
  redshift lensing sensitivity is relatively good, the volume probed
  is also quite limited.}
\label{james_plot}
\end{figure*}

Among group and cluster probes, X-rays are perhaps the cleanest and
the most complete selection method. Firstly, X-ray emission depends on
the square of the gas density and so X-rays pick up the cores of dense
structures more accurately than SZ and are less prone to projection
effects. Secondly, unlike optical techniques which rely on galaxy
properties, X-rays yield a complete sample of groups and clusters,
irrespective of their galaxy content. Finally, X-rays probe a wider
range in mass and redshift than shear maps which are fundamentally
limited by the shape of the lensing weight function. To illustrate the
magnitude of this effect, in the upper panel of Figure
\ref{james_plot} we show the expected lensing detection significance
of X-ray structures in COSMOS as a function of mass and redshift. The
lower panel in Figure \ref{james_plot} shows the comoving volume
probed by the survey per unit redshift. The theoretical lensing
detection significance is derived according to the method outlined in
\citet{Hamana:2004} assuming a smoothed COSMOS redshift distribution
and isolated NFW profiles truncated at the virial radius. A source
density of 66 galaxies/arcmin$^2$ and an average shape noise of
$\sigma_{\rm \tilde\gamma}=0.31$ is assumed. Lensing S/N curves are
based on fixed-scale Gaussian smoothing with a one arcminute smoothing
kernel (an optimal filter would pick up slightly more signal). Figure
\ref{james_plot} demonstrates that lensing alone cannot detect low
mass or high redshift objects. Instead one must resort to other
detection techniques such as X-rays. Note that although the low
redshift lensing sensitivity is relatively good, the volume probed is
also quite limited. It is also important to note that the depth of the
COSMOS data will probably exceed any near-future space-based
mission\footnote[4]{The fiducial depth of EUCLID and JDEM is $I\sim
  25.5$.}. In this respect, Figure \ref{james_plot} can be considered
as a realistic upper limit for lensing-based structure detection in
the near-future (for both ground and space-based observatories).

\subsection{Construction of the Group Catalog}
 
The group catalog used for this work is an improved version of the
catalog presented in \citet{Finoguenov:2007} (hereafter F07) obtained
by using additional {\sl XMM-Newton} and {\sl Chandra} data and by
applying a new procedure for the removal of point-sources (Finoguenov
et al., in prep). The group catalog contains a total of \ngroup
systems in the COSMOS ACS field, \ogroup of which are at $z<1$.

The group catalog is constructed from both the {\sl XMM-Newton} and
the {\sl Chandra} mosaics. All features with spatial extents below
16$^{\prime\prime}$ are removed before co-adding the two mosaics. The
combined mosaic is used to search for sources with spatial extents on
both 32$^{\prime\prime}$ and 64$^{\prime\prime}$ scales using the
wavelet decomposition technique described in
\citet{Vikhlinin:1998}. Both {\sl Chandra} and {\sl XMM-Newton} have a
spatial resolution that is better than 30$^{\prime\prime}$. The Full
Width Half Maximum (FWHM) of the PSF is approximately equal to
16$^{\prime\prime}$ for {\sl XMM-Newton}, and varies between
3$^{\prime\prime}$ and 8$^{\prime\prime}$ for {\sl Chandra}.

Total X-ray fluxes are obtained from the measured fluxes by assuming a
beta profile and by removing the flux that is due to embedded AGN
point-sources. Previous surveys have often assumed that all of the
X-ray flux within an extended source area is due to cluster emission
\citep[\eg][]{Bohringer:2004}. However, this assumption breaks down
for deep surveys such as COSMOS. In particular we have calculated that
on average, AGN contribute up to 30\% of the extended X-ray emission
in the 0.5--2 keV band, while for ROSAT All Sky Survey (RASS)
clusters, the estimated value is less than 2\%. Point-source removal
is thus necessary for COSMOS. However, the procedure that removes the
flux from point-sources also removes flux from cool-cores and as a
consequence, the total flux is underestimated. For comparison with
other work, it is important to correct for this accidental removal of
the flux from cool-cores. There have been claims that the cool-core
fraction is evolving with redshift
\citep[][]{Jeltema:2005,Vikhlinin:2007,Maughan:2007}. We have
therefore developed a method to correct for this effect directly from
the data by using the high-resolution {\sl Chandra} observations of
the COSMOS field which cover a contiguous area of 0.5 degrees$^2$ with
the best PSF \citep[3$^{\prime\prime}$,][]{Elvis:2009}, sufficient to
distinguish between cool-cores and AGN. Taking advantage of these
data, we compute a cool-core correction factor (noted $f_{CC}$) using
the following procedure. To begin with, groups inside the high
resolution {\sl Chandra} area are binned into the same nine bins as
used for the lensing analysis (see $\S$\ref{stacked_estimates}). Next,
wavelet scales below 4\arcsec\ are used to remove
point-sources. Finally, the {\sl Chandra} flux is stacked in each of
the nine bins and a 16\arcsec\ aperture is used to estimate the
average cool-core flux. The results are listed in Table 1 and range
from a 3\% to a 17\% flux correction.

Rest-frame luminosities are calculated from the total flux following
$L_{0.1-2.4 \rm keV} = 4\pi d_L^2 K(z,T) C_{\beta}(z,T) F_{\rm d}$,
where $K(z,T)$ is the K-correction and $C_{\beta}(z,T)$ is an
iterative correction factor (see F07). The uncertainty in $K(z,T)$
affects groups with luminosities below $10^{42} \lum$ and so only
concerns a few of the systems considered here. For the $L_{\rm X}-T$
relation, we have used $kT/keV=0.2+6\times10^{(\log (L_{\rm X}
  E(z)^{-1})-44.45)/2.1)}$ which introduces a break at group scales
\citep[as discussed in][]{Voit:2005} but which reproduces the
Markevitch (2002) result at cluster scales. Recent work by Pratt et
al. (2009) have derived a similar slope for the $L_{\rm X}-T$ relation
for clusters as Markevitch (2002). The behaviour of the $L_{\rm X}-T$
relation is not very well established at low temperatures, however
exploring the effects of a different $L_{\rm X}-T$ relations is beyond
the scope of this paper and is left for future work.

Quality flags (named hereafter ``{\sc xflag}'') are derived for the
entire group catalog based on visual inspection\footnote[5]{Visual
  inspection is performed on the XMM-Newton, ACS and Subaru data
  simultaneously.}. {\sc xflag=1} is assigned to groups with a single
optical counterpart and with a clear X-ray peak. These are mostly
high-significance X-ray detections (over 6$\sigma$) in zones free of
projection effects. The flag {\sc xflag=2} is assigned to systems for
which the extended X-ray emission is subject to projection effects but
for which the various projections can be disentangled. Systems with
questionable optical/IR counterparts are assigned {\sc xflag=3}. These
are primarily high-z ($z>1$) candidates that are not considered in
this work. {\sc xflag=4} indicates that there are several equally
possible optical counterparts and that the X-ray flux cannot be
disentangled for projection effects. {\sc xflag=5} is assigned to
systems for which the optical counterpart is uncertain and {\sc
  xflag=6} to identified extended emission not associated with galaxy
groups (these are mainly interacting galaxies and X-ray jets, Smolcic
et al., in prep.). {\sc xflag=7} is assigned to unidentified emission
and {\sc xflag=8} assigned to systems located in the masked-out
regions (edges of the survey and regions near bright stars). In this
work, we only consider systems with {\sc xflag}=1 or {\sc xflag}=2.

To ensure a high quality and robust lens catalog, in addition to the
X-ray quality flags, all systems have been visually inspected and
flagged for proximity to the edge of the ACS field, and contamination
by the presence of a bright foreground star which will affect both the
lensing measurements and the determination of the central group
galaxy. All systems with such flags were removed from the lens
catalog. In total, after all quality cuts, the lens catalog contains
\pgroup systems at $z \leq 1$. None of these quality cuts are expected
to bias the remaining sample but will improve the estimations of both
\lx and M$_{200}$.

\subsection{Redshift Determination}\label{redshift_determination}

The optical counterparts of X-ray sources are identified using a
sophisticated red-sequence method. The details of this technique are
presented in Finoguenov et al. (2009) -- only a brief outline is given
here. Along the line-of-sight of each X-ray source, the redshift range
$0<z<2.5$ is probed for potential red-sequence over-densities. For a
given redshift $z=z_{\rm RS}$ with $0<z_{\rm RS}<2.5$, galaxies are
selected within an aperture of 0.5Mpc (physical) from the center of
the X-ray emission such that $|z_{\rm phot}-z_{\rm RS}|<0.2$. The
apparent size of the aperture is defined in terms of a physical scale
and will vary with $z_{\rm RS}$. Weights are derived for all
galaxies according to their proximity to the center of the X-ray
emission and to the comparison of both their color and magnitude to a
model red-sequence at $z_{\rm RS}$. The red-sequence detection
significance is determined by applying the same procedure to random
COSMOS fields. If there are multiple red-sequence over-densities along
the line-of-sight, the one with the highest significance is
selected. Red-sequence redshifts are then refined using spectroscopic
information whenever possible.

Of the \pgroup systems at $z \leq 1$ (after the quality cuts), 81\%
have two or more spectroscopically confirmed members, 7\% have one
spectroscopically confirmed member, and 12\% have a redshift that is
determined solely by the red-sequence method. The average redshift
error for the group ensemble at $z \leq 1$ is estimated at $0.006$,
somewhat larger than the typical velocity dispersion of our groups
which is about 300 km s$^{-1}$.

\subsection{Center Determination}\label{center_determination}

Stacked weak lensing measurements require the identification of the
dark matter density peak. Centroid errors will lead to a smoothing of
the lensing signal on small scales and to an underestimate of halo
mass \citep[\eg\ see discussion in][]{Johnston:2007}. With the
exception of on-going, nearly equal mass ratio mergers, the centroid of
the X-ray emission should indicate where the potential well is
deepest. Halo centers are also often assumed to contain central
galaxies (CGs) which can be used as good tracers on condition that
they can be correctly identified.

The wavelet-reconstructed X-ray image is analysed with SExtractor
\citep{Bertin:1996} to determine the centers and 2d shapes of the
extended X-ray emission. The accuracy of the determination of the
X-ray center is higher for {\sc xflag}=1 systems than for {\sc
  xflag}=2 systems which are somewhat affected by projection
effects. The maximum uncertainty of the peak for {\sc xflag}=2 systems
is determined by the size of the wavelet scale, which is
$32^{\prime\prime}$. Although the X-ray centroid is not precise enough
to be used directly in most cases (of \pgroup system that remain after
the quality cuts described in previous sections, 55 have {\sc xflag}=1
and 72 have {\sc xflag}=2), it is precise enough to be used as a
strong prior on the location of the CG. Indeed, the maximum
uncertainty in the X-ray centroid is $32^{\prime\prime}$ (193
$h_{72}^{-1}$ kpc at z=0.5), however, most systems have a smaller
positional uncertainty than this. For comparison
purposes, the average projected radial offset for mis-identified CGs
in MaxBCG is larger than 600 $h_{72}^{-1}$kpc \citep[Figure
5,][]{Johnston:2007} (note that their distances must be converted to
our assumed value of $H_0$).

In many previous studies, Brightest Cluster Galaxies (BCGs) have been
associated with the CGs of group and cluster halos. Given the
ambiguity in the choice of the filter in which BCG galaxies should be
taken as ``brightest'' as well as the sensitivity of optical
luminosity to recent star-formation, we prefer the utilization of the
Most Massive Central Galaxy (MMCG) located near the peak X-ray
emission (where massive refers to stellar mass). We assume that the
MMCG can be used to trace the center of the dark matter halos of
groups and clusters. An automatic algorithm was developed to identify
MMCGs. Briefly, for each X-ray group, a broad group member selection
is made by selecting galaxies within 800kpc of the peak X-ray
emission such that $|z_{phot}-z_{group}|< 0.03 \times
(1+z_{group})$. Next, galaxies are rank-ordered by their stellar mass
and weighted by the proximity to the peak X-ray emission -- the MMCG
is the galaxy with the highest rank. The results were visually
inspected and divided into three categories:

\begin{enumerate}
\item {\sc cg-type}=1: the CG is visually obvious (for the most part,
  a dominant early type galaxy with an extended envelope) and the
  algorithm has selected the correct galaxy.
\item {\sc cg-type}=2: there is a visual ambiguity in the CG selection
  but we estimate that the algorithm has selected a galaxy that has a
  $50\%$ chance of being the CG.
\item {\sc cg-type}=3: the visual identification is highly ambiguous
  or there is some other problem that prevents the identification of
  the CG.
\end{enumerate}

In this study, we only consider {\sc cg-type}=1 and {\sc cg-type}=2
systems. In combination with the quality cuts described previously,
these cuts leave a total of \qgroup groups and clusters at $z \leq 1$
(95 of which are {\sc cg-type}=1 and 23 are {\sc cg-type}=2). The
details of the MMCG selection as well as tests regarding centering
uncertainties will be presented in a forthcoming paper (George et al.,
in prep).

In terms of stacked weak-lensing, there are two mis-centering effects
to be taken into consideration. The first is that the location of CGs
could be poor tracers of the actual centers of their dark matter
halos. The second is that the CGs could be mis-identified. In a
similar fashion to the maxBCG studies
\citep[][]{Sheldon:2007,Johnston:2007}, we neglect the
former. However, our analysis differs from the maxBCG studies with
respect to the latter. \citet{Johnston:2007} assume a CG
mis-identification fraction of $\sim$30\% and apply a mis-centering
kernel in their analysis to account for this effect. In this study, we
assume that our X-ray prior combined with thorough visual checks,
allows us to correctly identify the CG for a majority of our systems
and that when a CG is mis-identified, the projected radial offset from
the dark matter center is not large. Testing this assumption in
further detail will be the focus of a subsequent paper (George et al.,
in prep). Nonetheless, in $\S$\ref{syst_error} we have also
demonstrated that restricting our analysis to {\sc cg-type}=1 systems
does not affect our results, indicating that errors due to
mi-centering are probably not a dominant effect for this work.

We also note that our CG selection is based on stellar mass and is
insensitive to color, hence we avoid the problem of blue core BCGs
which represent about $25\%$ of the BCG population according to
\citet{Bildfell:2008}. The excess blue light in these systems can lead
to a typical offset from the red sequence of 0.5 to 1.0 mag in (g' -
r') which could lead to their rejection by red-sequence type methods.


\section{Functional Form of $\mathrm{M}-\mathrm{L}_{\rm X}$}\label{lx_m_relation}

In this section we present previously published results and overview
the various assumptions that are made regarding the functional form of
$\rm{M} - \rm{L}_{\rm X}$. We also explain how the scatter between
mass and luminosity can cause subtle differences between the study of
M as a function of \lx (the ``\mlx relation'') and the study of \lx as
a function of M (the ``\lxm relation'').

Previous measurements of the \lxm relation based on X-ray data
\citep[][]{Reiprich:2002,Allen:2003,Popesso:2005,Chen:2007,Pratt:2008,Zhang:2008,Vikhlinin:2009}
as well as the the \mlx relation derived with lensing
\citep[][]{Hoekstra:2007a,Bardeau:2007,Rykoff:2008} are by-and-large
consistent with a power-law, but with a slope and amplitude that
differ from the self-similar prediction of $M \propto L_{\rm
  X}^{3/4}$. In contrast, the evolution of the \lxm relation is still
under much debate with certain authors finding that the \lxm relation
evolves in a self-similar fashion
\citep[][]{Lumb:2004,Arnaud:2005,Kotov:2005,Maughan:2007} while others
do not \citep[][]{Ettori:2004}.

In addition to the shape and evolution of the mean \lxm relation,
astrophysical processes are expected to induce scatter in \lx at fixed
mass which is important to take into consideration. In the absence of
strong observational or theoretical guidance for the form and
magnitude of this scatter \citep[although
see][]{Reiprich:2002,Maughan:2007}, it is common to adopt a stochastic
model where $P(L_{\rm X}|M)$ is a log-normal probability distribution
function (hereafter PDF) with a mean log-luminosity that follows
$\langle \ln L_{\rm X} \rangle \propto \beta \ln M$ and with a
constant log-normal scatter noted $\sigma_{\ln L_{\rm X}}$
\citep[][]{Stanek:2006, Rozo:2008, Rykoff:2008, Vikhlinin:2009}. In
this particular case, and under the further condition that the halo
mass function is a power-law, it can be demonstrated that $P(M|L_{\rm
  X})$ is also a log-normal probability function with a dispersion in
mass equal to $\sigma_{\ln M}=\sigma_{\ln L_{\rm X}}/\beta$ and a mean
log-mass that follows $\langle \ln M \rangle \propto \alpha \ln L_{\rm
  X}$ with $\alpha=1/\beta$ (see Appendix for further details). As a
consequence, the slopes of the \lxm and the \mlx relation can be
compared quite simply but the comparison of the normalization will
depend on the halo mass function.

The difference in the normalization of $P(L_{\rm X}|M)$ and
$P(M|L_{\rm X})$ can be seen as a form of extended Malmquist
bias. This is not Malmquist bias is the classical sense because it
will occur in any survey, independently of the flux limit. The
equations for this bias are derived in the Appendix. In general, X-ray
astronomers commonly employ $P(L_{\rm X}|M)$ \citep[][]{Stanek:2006,
  Vikhlinin:2009} whereas lensing more naturally derives $P(M|L_{\rm
  X})$\citep[][]{Rozo:2008, Rykoff:2008} so care must be taken when
comparing the two.

In reality, the slope of the mass function varies with both mass and
redshift and as a consequence, $\alpha=1/\beta$ no longer holds when
slopes are derived over a large range in masses. Corrections for this
effect are derived in the Appendix.

\begin{figure*}[htb]
\centerline{\includegraphics[scale=0.53]{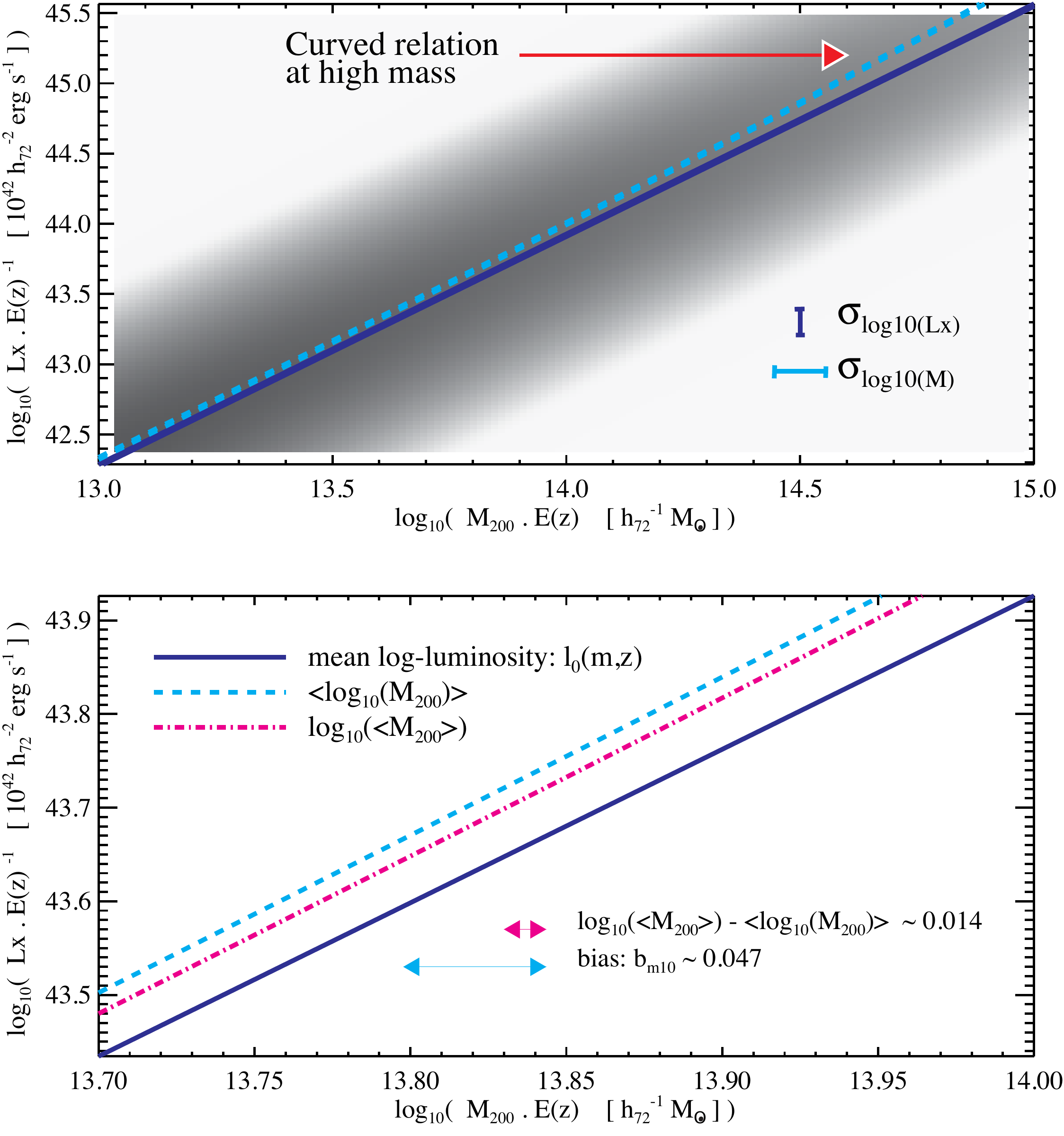}}
\caption{Illustration of the effects of biases in the \lxm and the
  \mlx relations at $z=0.2$. The conditional probability distribution
  of the luminosity given the mass, $P(\mathrm{L}_X|\mathrm{M},z)$, is
  assumed to be log-normal with a mean log-luminosity that follows a
  power-law scaling relation (blue solid line) with a slope of
  $1.63=1/0.61$ and with a scatter of $\sigma_{\ln l}=0.4$. A
  simulated ensemble of groups and clusters of galaxies is created for
  which the number densities per mass follow the
  \protect\citet{Tinker:2008} halo mass function and for which X-ray
  luminosities are attributed according to
  $P(\mathrm{L}_X|\mathrm{M},z)$. The grey shading in the upper panel
  is proportional to the log of the number density of groups and
  clusters in the luminosity-mass plane (arbitrary normalization). The
  mean log-mass follows the cyan dashed line and the log of the
  mean-mass ($\log_{10}\langle M_{200} \rangle$, measured by lensing)
  is shown by the dash-dot magenta line. There is a mass-dependent
  bias between the cyan and the dark blue line equal to
  $b_{m}=\sigma_{m}^2(\gamma-1)$ where $\gamma$ is the slope of the
  mass function (see Appendix). As a results of this bias, there is a
  slight curvature in both the cyan and the magenta lines at high halo
  masses.}
\label{bias_illustration}
\end{figure*}

Stacked weak lensing yields a measurement of the arithmetic mean of
the surface mass density contrast, $\langle \Delta\Sigma(r)
\rangle$. When the data are binned according to a well chosen proxy
and over a narrow redshift range (to avoid smearing the profiles due
to evolution in the mass-concentration relation for example), the mass
derived by fitting $\langle \Delta\Sigma(r) \rangle$ will be close to
to the arithmetic mean of the stacked ensemble $\langle M_{200}
\rangle$. For this reason, we select narrow redshift bins that further
enable us to assume that $\langle M_{200}.E(z) \rangle \sim \langle
M_{200}\rangle E(\langle z \rangle)$. Note that if the PDF of the mass
at fixed luminosity is log-normal, then $\langle M_{200} \rangle$ will
be {\em different} from the peak of the PDF. Indeed, in this case, the
peak is traced by the median, not the arithmetic mean. If the scatter
of $P(M|L)$ is known, then the correspondence between the two is given
by $\langle M_{200} \rangle=\exp (\langle \ln (M_{200})
\rangle+\sigma_{\ln M}^2/2)$. Figure \ref{bias_illustration}
illustrates the various issues outlined above and for which more
detailed calculations are presented in the Appendix.

Estimates for the scatter\footnote[7]{Scatter is quoted as the
  standard deviation of the natural logarithm of the mass at fixed
  \lx.} in the \mlx relation vary from $\sigma_{\ln M} \sim 0.2$ to
$\sigma_{\ln M} \sim 0.3$ and can depend exactly on how \lx is
measured. The lowest scatter is obtained with cool-core excision
techniques \citep[][]{Stanek:2006, Maughan:2007, Pratt:2008}. For
distant clusters, the cluster core region can become smaller than the
observed PSF and so cool-core excision becomes infeasible. However,
there have been suggestions that the cool-core fraction is low at
$z>0.5$ \citep[][]{Vikhlinin:2007}, perhaps making excision
unnecessary at higher redshifts. In addition, \citet[][]{Maughan:2007}
have suggested that $\sigma_{\ln M}$ is reasonably small even for
survey quality data. In any case, although the scatter in mass at
fixed luminosity is still poorly constrained (especially at high
redshift), by most estimates it is smaller than the scatter in mass at
fixed richness, even for the best richness estimators \citep[\eg][find
$\sigma_{\ln M |\lambda}=0.45$]{Rozo:2008}.

Given the various considerations discussed above, we adopt a power-law
form for the mean relation between mass and luminosity with a redshift
evolution that follows self-similarity:

\begin{equation}
  \frac{\langle  M_{200}~E(z) \rangle}{M_0} = A \left( \frac{\langle L_{X}~E(z)^{-1}\rangle }{L_{X,0}} \right)^{\alpha}
\label{eq:lx_m}
\end{equation}

\noindent where $M_0=10^{13.7} \mass$ and
$L_{X,0}=10^{42.7} \lum$. Deviations from self-similar evolution are
tested for in $\S$\ref{evolution}.

To begin with, we derive the relationship between the mean mass
$\langle M_{200}\rangle E(\langle z \rangle)$ and the mean luminosity
$\langle L_{X}~E(z)^{-1}\rangle $ using only the COSMOS data. This
relation is referred to as $\mathcal{R}1_{M-L_{X}}$. Next, we combine
the COSMOS results with previously published cluster data to improve
measurements of the slope $\alpha$. This combined relation is referred
to as $\mathcal{R}2_{M-L_{X}}$.


\section{Results}\label{results}

In this section we present our lensing measurements as well as the
\mlx relations $\mathcal{R}1_{M-L_{X}}$ and
$\mathcal{R}2_{M-L_{X}}$. We also test for additional redshift
evolution in the \mlx relation beyond that predicted by
self-similarity.

\subsection{Stacked estimates of $\Delta\Sigma$ and $M_{200}$}\label{stacked_estimates}

\begin{figure*}[htb]
\centerline{\includegraphics[scale=0.43]{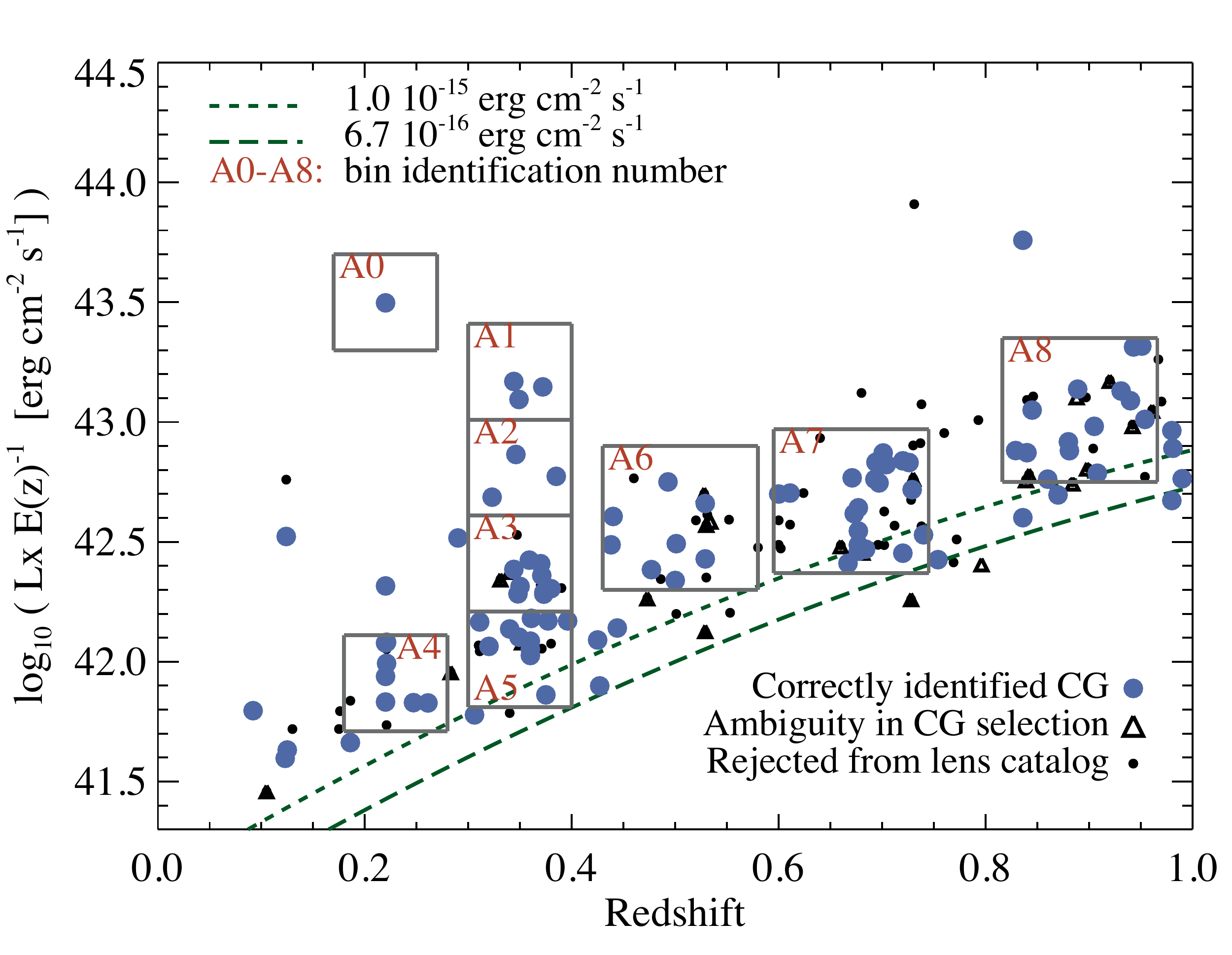}}
\caption{COSMOS groups within the ACS field as a function of redshift
  and $L_{\rm X}E(z)^{-1}$, binned into nine sub-samples labeled
  $A_{0}$ through $A_{8}$. Each bin is designed to encompass a narrow
  range in both redshift and $L_{\rm X} E(z)^{-1}$. The 4$\sigma$
  sensitivity limits are shown by the green lines: 96$\%$ of the ACS
  field is covered to a sensitivity limit of $1.0~10^{-15}$ erg
  cm$^{-2}$ s$^{-1}$ where the sensitivity limits have been derived
  using a wavelet sensitivity map on group scales (green dotted line),
  and 52$\%$ of the ACS field is covered to a deeper limit of
  $6.7~10^{-16}$ erg cm$^{-2}$ s$^{-1}$ (green dashed line). Small
  black circles represent systems that were rejected from the lens
  catalog. Blue filled circles indicate systems for which the central
  galaxy (CG) identification is certain whereas black triangles show
  systems for which the identification is more ambiguous.}
\label{bining_scheme}
\end{figure*}

The data are divided into nine bins labeled $A_0$ through $A_8$ (see
Figure \ref{bining_scheme} and Table \ref{lx_bins}). The bins are
selected to encompass a narrow range in redshift and $L_X E(z)^{-1}$
so as to avoid smearing out the signal due to evolution in the
mass-concentration relation. For each bin, the stacked weak lensing
signal is calculated according to the method outlined in
$\S$\ref{lensing_theory}.

\begin{deluxetable*}{lccccccccccc}
  \tabletypesize{\scriptsize} \tablecolumns{10} \tablecaption{Various properties for each of the nine bins\label{lx_bins}} \tablewidth{0pt} 
\startdata
\hline 
\hline 
\\  [-1.5ex]
Bin ID & N$_{LENS}$ & $\langle {L_{X}.E(z)^{-1}} \rangle$ & $f_{\rm CC}$\tablenotemark{a}  & $\langle {M}_{200} \rangle$ & $\langle z \rangle$ & E($\langle z \rangle $) & $L_{Z}$ & $f_{bias}$ & $f_{boost}$ & $f_{bias} \times f_{boost}$ & \\ [1ex]
       &          &  (10$^{42}$ $h_{72}^{-2}$ erg s$^{-1}$) & & (10$^{13}$ $h_{72}^{-1}$ M$_{\odot}$)   &  & &(10$^{-4}$ $h_{72}^{-2}$ pc$^2$ M$_{\odot}^{-2}$)& & & &\\ [1ex]
\hline\\  [-1.5ex]
A0 & 1  & 31.14 $\pm$0.49  & 1.0  & 14.9$^{+7.1}_{-4.8}$ & 0.22 & 1.10 & 3.16 & 1.01 & 1.00 &  1.01 \\ 
A1 & 3  & 13.75 $\pm$0.50  & 1.0  & 8.2$^{+3.3}_{-2.3}$  & 0.36 & 1.18 & 3.78 & 1.02 & 1.00 &  1.03 \\ 
A2 & 3  & 6.04  $\pm$0.28  & 1.17 & 9.9$^{+3.3}_{-2.5}$  & 0.35 & 1.17 & 3.77 & 1.01 & 1.00 &  1.02 \\ 
A3 & 11 & 2.21  $\pm$0.15  & 1.08 & 3.2$^{+1.3}_{-0.9}$  & 0.36 & 1.18 & 3.77 & 1.02 & 1.00 &  1.02 \\ 
A4 & 7  & 0.90  $\pm$0.06  & 1.03 & 2.1$^{+1.5}_{-0.8}$  & 0.23 & 1.11 & 3.20 & 1.01 & 1.00 &  1.01 \\ 
A5 & 13 & 1.24  $\pm$0.09  & 1.05 & 1.1$^{+0.8}_{-0.4}$  & 0.35 & 1.18 & 3.77 & 1.01 & 1.00 &  1.02 \\ 
A6 & 11 & 3.65  $\pm$0.21  & 1.17 & 3.3$^{+1.7}_{-1.1}$  & 0.50 & 1.27 & 3.96 & 1.03 & 1.01 &  1.04 \\ 
A7 & 23 & 4.72  $\pm$0.24  & 1.13 & 2.6$^{+1.1}_{-0.7}$  & 0.69 & 1.41 & 3.84 & 1.04 & 1.02 &  1.06 \\ 
A8 & 21 & 10.51 $\pm$0.50  & 1.14 & 7.6$^{+2.3}_{-1.8}$  & 0.90 & 1.58 & 3.65 & 1.03 & 1.05 &  1.08 \\ 
\enddata
\tablenotetext{a}{The cool-core correction factor that is applied to column 3 (see description in $\S$2.1).}
\end{deluxetable*}

We calculate the weak lensing signal from \srmin to \srmax in
logarithmically spaced radial bins of 0.26 dex. A weak lensing signal
is detected all the way to $4$Mpc, allowing us to probe the full
extent of the one-halo term. The results are fit with the parametric
model given by Equation \ref{ds_model}.

We use a Markov chain Monte Carlo (MCMC) method to fit the
$\Delta\Sigma$ profiles. The MCMC routine uses the Metropolis-Hastings
algorithm with a Gaussian transfer function. The total number of steps
is $30000$ and a burn-in period of $500$ is discarded. Bins with less
than 15 background sources in total are excluded from the fit.

\begin{figure*}[htb]
\centerline{\includegraphics[scale=0.42]{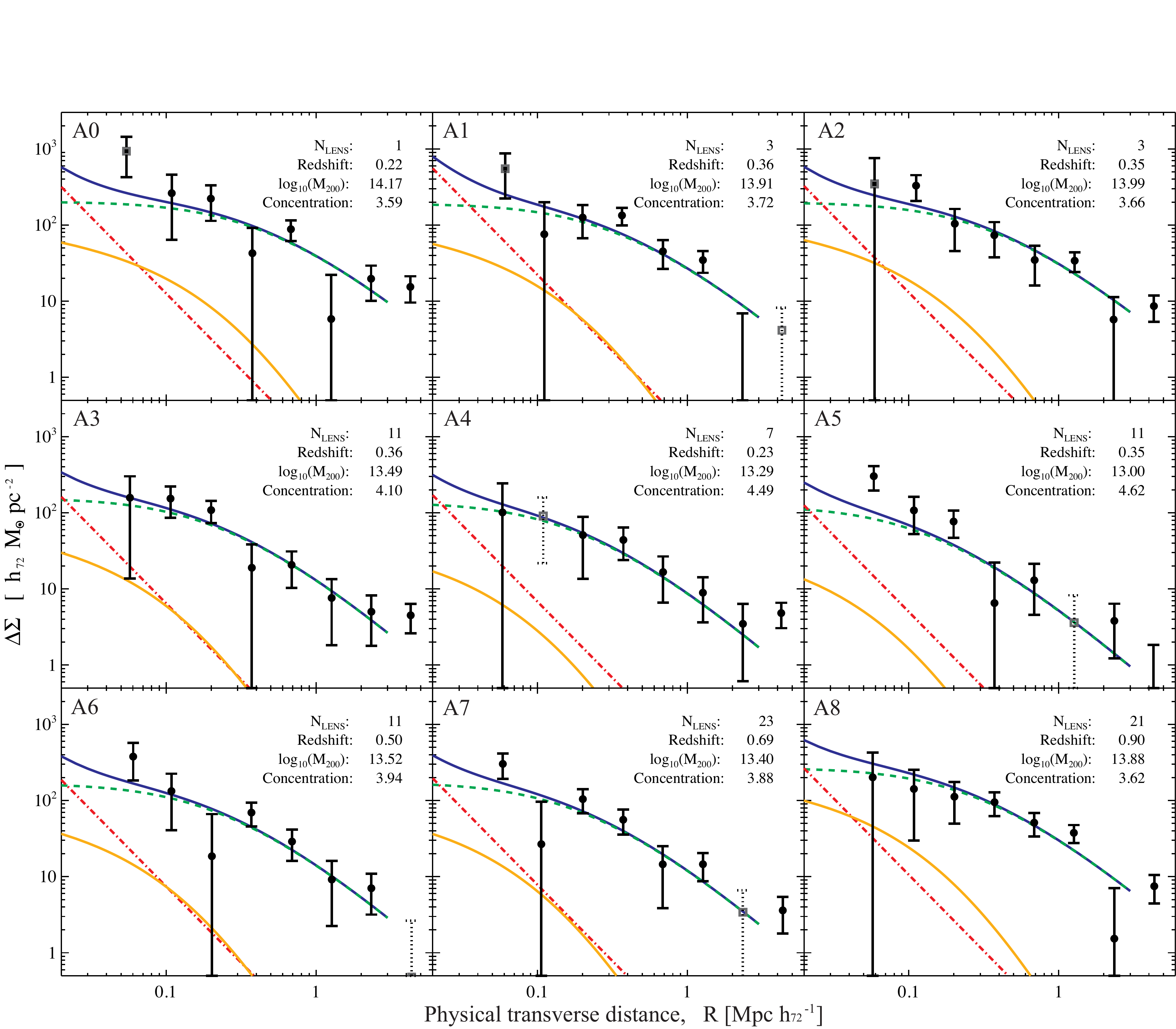}}
\caption{Stacked weak lensing profiles of COSMOS groups for nine bins
  that each span a narrow range in redshift and $L_X E(z)^{-1}$. From
  left to right and top to bottom we show the bins $A_0$ through
  $A_8$. A weak lensing signal is detected all the way to $4$Mpc,
  allowing us to probe the lens density profiles well beyond the
  virial radius. The solid blue curve shows our fit to the data which
  is the sum of a baryonic term (red dash-dot), an NFW profile (green
  dash-dash), and a second order weak shear correction term (orange
  dash-dot-dot, see $\S$ \ref{non_weak_shear}). Grey points are
  negative data-points and bins with less than 15 source galaxies.}
\label{wl_signal}
\end{figure*}

The results from the stacked analysis and the fits to the profiles are
shown in Figure \ref{wl_signal}. At the smallest radii that we probe,
the stellar mass of the central galaxy plays a minor role in the
lensing signal but we have added it for consistency. The scales that
we probe are dominated by the signal due to the dark matter halo
associated with the groups. Our estimates for $\langle M_{200}
\rangle$ are given in Table \ref{lx_bins}.

\subsection{Measurement of  $\mathcal{R}1_{M-L_{X}}$}\label{lxm_relation1}

\begin{figure}
\centerline{\includegraphics[scale=0.47]{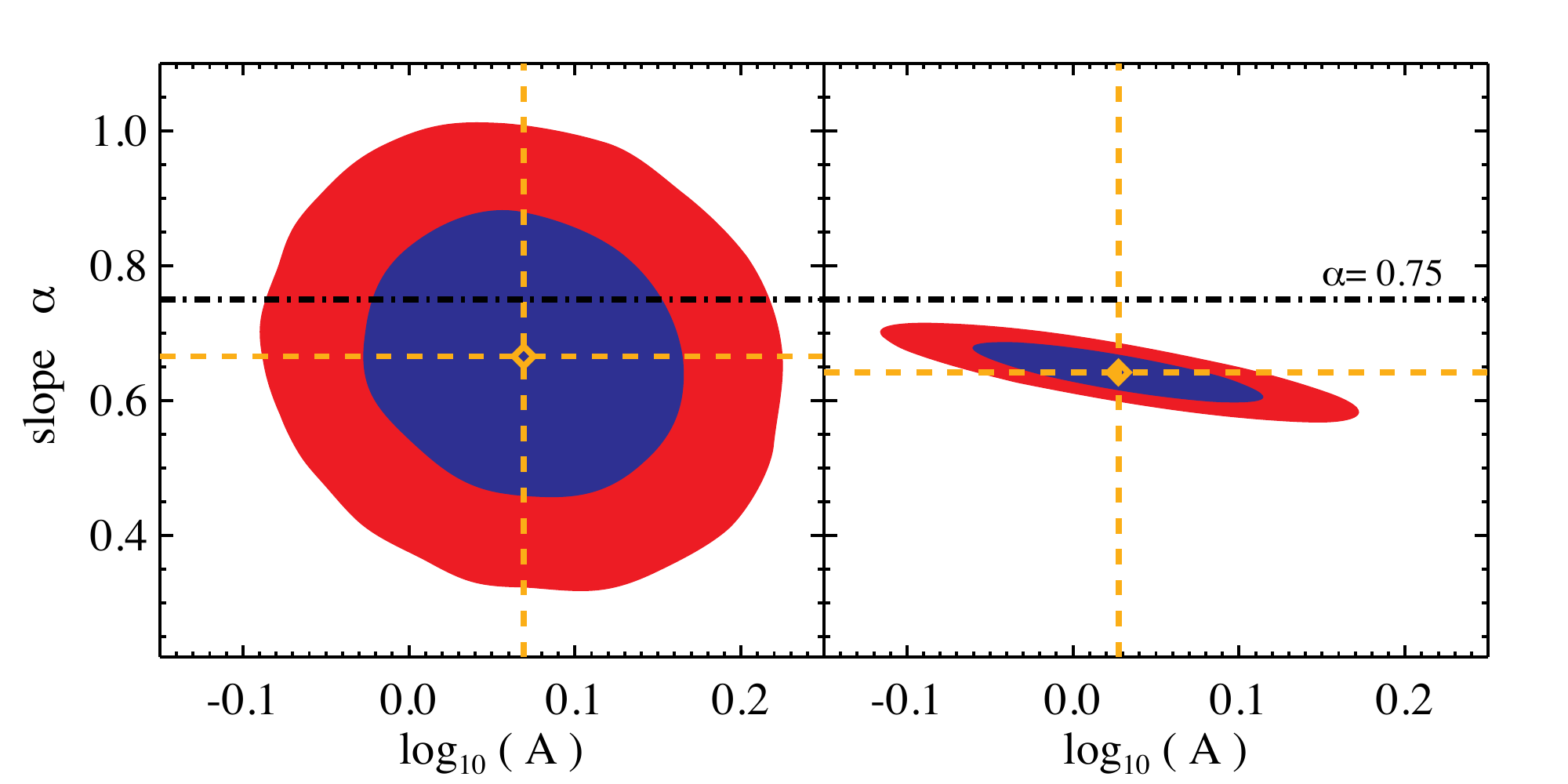}}
\caption{Left panel: marginal posterior distribution for the
  parameters of $\mathcal{R}1_{M-L_{X}}$ (COSMOS data only). The blue
  shaded region denotes the 68$\%$ ($1 \sigma$) confidence region and
  the red shaded region denotes the 95$\%$ ($2 \sigma$) confidence
  region. The dash-dot line indicates the self-similar prediction for
  the slope of the \mlx relation. Right panel: marginal posterior
  distribution for the combination of the COSMOS group data and
  cluster data from \protect\citet{Hoekstra:2007}. The large dynamic
  mass range created by the combination of group and cluster data
  enables a more accurate determination of the slope of the \mlx
  relation.\vspace{0.05 in}}
\label{lxm_a_alpha}
\end{figure}

Fitting only the low redshift COSMOS data (from A0 to A6,
$0.2<z<0.5$), we obtain the best fit parameters $\log_{10}(A)=0.068
\pm 0.063$ and \cosalpha\ (see Figure \ref{lxm_a_alpha}). The cited
errors are statistical only. The effects of systematic errors are
explored in $\S$\ref{syst_error} and are estimated to be lower than
the statistical uncertainty. The cool-core correction factor that we
apply does not strongly affect these results. Without the cool-core
correction, we obtain $\log_{10}(A)=0.09 \pm 0.062$ and $\alpha=0.64
\pm 0.14$.

\begin{figure*}[htb]
\centerline{\includegraphics[scale=0.75]{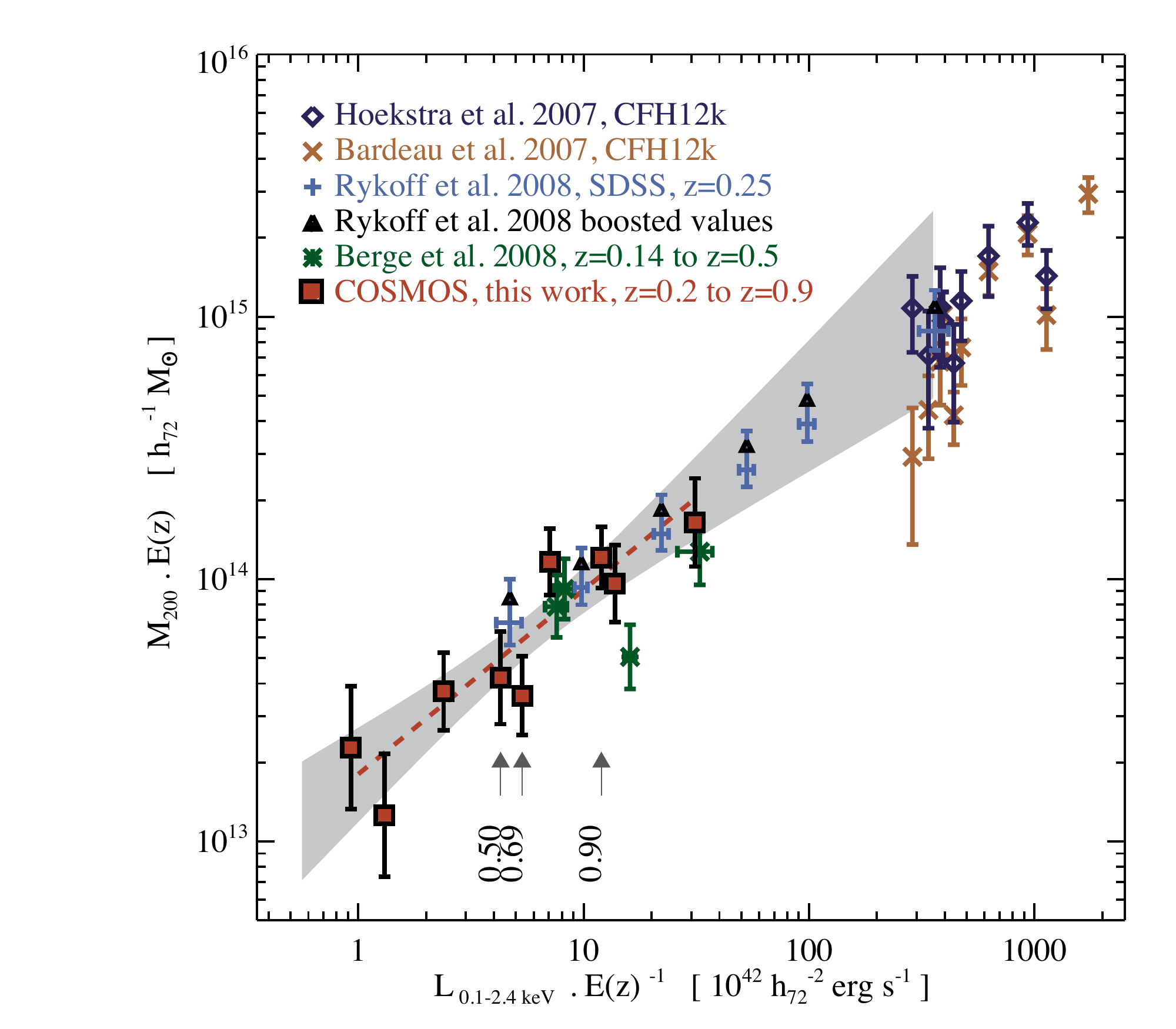}}
\caption{The COSMOS \mlx relation. Dark blue diamonds show
  individually detected clusters from H07 with updated masses from
  Madhavi et al. 2008. Sienna cross symbols show data points from
  BA07. Light blue plus symbols represent the R08 results from a
  stacked analysis in the SDSS and black diamonds take into account a
  recent correction to these masses due to a new calibration of the
  source distribution. The upper error bars have been adjusted to
  account for the redshift uncertainty. Green asterisks show four data
  points at intermediate masses from BE08. Finally, the red squares
  depict our COSMOS results which extend previous results to lower
  masses and to higher redshifts. Three arrows highlight the highest
  redshift COSMOS data points (bins $A_6, A_7$ and $A_8$). The grey
  shaded region shows the upper and lower envelope of the ensemble of
  lines with a slope and intercept that lie within the 68 percent
  confidence region of $\mathcal{R}1_{M-L_{X}}$.}
\label{lx_m}
\end{figure*}

Figure \ref{lx_m} compares the COSMOS $\mathcal{R}1_{M-L_{X}}$
relation to four other lensing-based measurements: the SDSS results
from \citet{Rykoff:2008} (hereafter R08), four groups from
\citet{Berge:2008} (hereafter BE08), cluster data from
\citet{Hoekstra:2007} (hereafter H07) with masses updated in
\citet{Mahdavi:2008}, and cluster data from \citet{Bardeau:2007}
(hereafter BA07). All data points have been normalized to $H_0=72$
$h_{72}$ km~s$^{-1}$~Mpc$^{-1}$ and scaled by $E(z)$. The cluster data
points from BA07 and the H07 have been selected on the basis that
their lensing analysis extends to the virial radius, allowing them to
derive mass estimates that are comparable to ours. Each of these four
independent studies probes a distinct redshift and mass
scale. Nevertheless, the ensemble of data points displays a remarkable
trend that spans over three orders of magnitude in ${\rm L}_{\rm X}
E(z)^{-1}$ and two orders of magnitude in ${\rm M}_{\rm 200}E(z)$. The
varying degrees of scatter seen between the different data-sets is due
to the fact that some results are direct detections of individual
clusters (\eg\ H07, BA07 and BE08) while other data points have been
stacked (\eg\ R08). The COSMOS data points are scattered about the
mean relation because each bin contains a relatively small number of
groups (tens of groups as opposed to hundreds in R08). We will now
briefly describe each of these data sets. Further discussion of the
agreement between various results is presented in
$\S$\ref{discussion}.

The R08 data points (light blue, plus signs) are taken from Table 1 of
their paper.  \lx and \m have been normalized to our adopted value of
$H_0$ and scaled with the $E(z)$ factor at the quoted redshift of
$z=0.25$. \citet{Mandelbaum:2008} have shown that the masses published
by \citet{Johnston:2007} which have been used in the R08 analysis,
must be boosted by a factor of $1.18^{1.4} = 1.25$ when the SDSS
source distribution is calibrated against zCOSMOS spectroscopic
redshifts. We boost the R08 data points by a factor of 1.24 (a revised
version of the published correction, R. Mandelbaum, priv. comm.) and
the results are shown by the small black triangles in Figure
\ref{lx_m}. The upper error bars of the R08 data points are increased
in order to reflect this correction. Note that because both \m and \lx
have been derived via stacking methods, the results of R08 will depend
on the covariance between $L_{X}$ and richness (noted $N_{200}$, see
R08) at fixed mass: the slope of their relation will change depending
on the correlation between these two parameters. This is not an issue
for COSMOS where the stacking is performed directly on \lx instead of
using an intermediate variable such as N$_{200}$.

The masses for the BE08 data points (dark green, asterisk signs) are
taken from their paper and \lx has been provided by F. Pacaud
(priv. comm.).

The masses for the BA07 data points (sienna, cross signs) are taken
from their paper. X-ray luminosities are derived from the XMM LOCUSS
survey using the same X-ray data as in \citet{Zhang:2008} but by
integrating the flux out to the truncation radius of $2.5
r_{500}^{Y_{\rm X}, wl}$. Note that $r_{500}^{Y_{\rm X}, wl}$ is given
in $\S$6.2.3 in \citet{Zhang:2008}. The imaging data for the BA07
analysis is based on ground-based wide-field imaging obtained with the
CFH12k camera on the Canada France Hawaii Telescope (CFHT). One
cluster in the BA07 paper did not have an XMM LOCUSS luminosity and
has been discarded (namely A2219).

The H07 data points (dark blue diamonds) have been taken from column 8
of Table 1 in \citet{Mahdavi:2008}. Luminosities are derived from the
LOCUSS survey using the methodology described in the previous
paragraph. The overlap between the surveys is the following set of
clusters: A68, A209, A267, A383, A963, A1689, A1763, A2218, and
A2390. Note that BA07 and H07 have studied a {\em common set of
  clusters} using the same imaging data (CFH12k camera). Differences
is cluster mass estimates between BA07 and H07 are probably due to
dissimilar data analysis techniques (such as assumptions regarding the
mass-concentration relation for example).

\subsection{Measurement of  $\mathcal{R}2_{M-L_{X}}$}\label{lxm_relation2}

In this section, we perform a joint fit between the COSMOS data and
cluster data from H07 and BA07. The high redshift COSMOS data points
(bins $A_7$ and $A_8$) are excluded from this fit so that all data
points are at a comparable redshift ($z\sim 0.3$). The joint fit
COSMOS/H07 yields $\log_{10}(A)=0.03 \pm 0.06$ and \henkalpha. The
joint fit COSMOS/B07 also yields $\log_{10}(A)=0.03 \pm 0.06$ and
$\alpha=0.64 \pm 0.03$. These results raise several points of
interest. Firstly, the fit to the COSMOS data alone gives a very
similar relation to the fit when the cluster data is added. This
suggests that the \mlx relation is invariant from group to cluster
scales with no detected break at group scales. Secondly, the
combination of group and cluster data creates a large dynamic mass
range which allows for a $5\%$ (statistical error) determination of
the slope of the \mlx relation. Finally, despite the systematic
differences in the mass estimates of H07 and BA07, the combined
relations with COSMOS are identical. The addition of group data has
significantly reduced the impact of cluster mass uncertainties on the
measured \mlx relation.

\subsection{Redshift Evolution in the \mlx relation}\label{evolution}

In this section we test for redshift evolution in the \mlx relation by
adding an additional redshift dependent term to our previously assumed
\mlx relation as follows:

\begin{equation}
  \frac{\langle M_{200}~E(z)\rangle }{M_0} = A \left( \frac{\langle L_{X}~E(z)^{-1}\rangle }{L_{X,0}} \right)^{\alpha}E(z)^{\gamma}
\label{eq:lx_m2}
\end{equation}

\begin{equation}
  \frac{\langle M_{200}~E(z)\rangle }{M_0} = A \left( \frac{\langle L_{X}~E(z)^{-1}\rangle }{L_{X,0}} \right)^{\alpha}(1+z)^{\delta}
\label{eq:lx_m3}
\end{equation}

Fitting the COSMOS data alone with Equation \ref{eq:lx_m2} yields
$\log_{10}(A)=0.05 \pm 0.1$, $\alpha=0.70 \pm 0.13$, and $\gamma=-0.07
\pm 0.9$. Equation \ref{eq:lx_m3} yields $\log_{10}(A)=0.07 \pm 0.14$,
$\alpha=0.70 \pm 0.13$, and $\delta=-0.14 \pm 0.8$. In both cases, the
COSMOS data is consistent with self-similar redshift evolution, to the
level that can be probed with these data. Additional cluster and group
data with weak-lensing masses at $0.6<z<1.0$ would be required in
order to constrain the redshift evolution in the \mlx relation to
higher precision.

\section{Assessment of Systematics Errors}\label{syst_error}

There are three potential causes of systematic errors in this work:
errors in the photometric redshifts, mis-centering, and uncertainty in
the mass-concentration relation. As demonstrated below, we have found
that the systematic errors associated with each of these effects are
below the statistical error, at least to the extent that this can be
tested for with the current data.

\subsection{The impact of photometric redshift errors}

The effects of redshift errors on group-galaxy lensing signals can be
broadly categorized as follows: a) uncertainties in the redshifts of
the lenses will smear out the signal and affect the derivation of
$\Sigma_{\rm crit}$, b) errors in the mean source redshift
distribution will introduce a bias in the normalization of the overall
signal, c) improper lens-source separation will lead to a dilution of
the signal and will subject the signal to the effects of intrinsic
alignments, d) and catastrophic errors can also dilute the signal. In
this work, we neglect (a) on the basis that the redshifts of the
groups are well determined (81\% have two or more spectroscopically
confirmed members).

To reduce the effects of (c) and (d), we use the full redshift PDF
which is derived by the photoz code LePhare for each source galaxy
\citep[][]{Ilbert:2009}. The main peak in the PDF, {\sc zp\_max},
represents the most probable redshift. When present, a secondary peak
in the PDF is noted {\sc zp\_sec}. The galaxy population with double
peaked PDFs is expected to contain a large fraction of catastrophic
errors \citep[roughly 40\%-50\%,][]{Ilbert:2006,Ilbert:2009}. In this
work, we eliminate sources with a double peaked PDF in order to
minimize the effects of signal dilution caused by catastrophic
errors. Making this photoz cut leads to a decreased background number
density of 34 galaxies/arcmin$^2$.

The redshift PDF information is also used to improve the lens-source
separation by using the lower $68\%$ confidence bound on the source
redshift. Sources are selected such that $z_S-z_L>
\sigma_{68\%}(z_S)$. It is also important to note that group-galaxy
lensing signals are most sensitive to redshift errors when $z_S$ is
only slightly larger then $z_L$ (see Figure 8). For this reason, in
addition to the previous cut, we also implement a fixed cut such that
$z_S-z_L> \delta z$, where $\delta z$ is defined below in the following two
schemes:

\begin{itemize}
\item $\mathcal{S}1$: Only sources with a single peaked PDF are used,
  $z_S-z_L> \sigma_{68\%}(z_S)$, and $z_S-z_L> 0.1$ (the default
  scheme used throughout this paper).
\item $\mathcal{S}2$: Only sources with a single peaked PDF are used,
  $z_S-z_L> \sigma_{68\%}(z_S)$, and $z_S-z_L> 0.2$.
\end{itemize}

\begin{figure*}[htb]
\centerline{\includegraphics[scale=0.47]{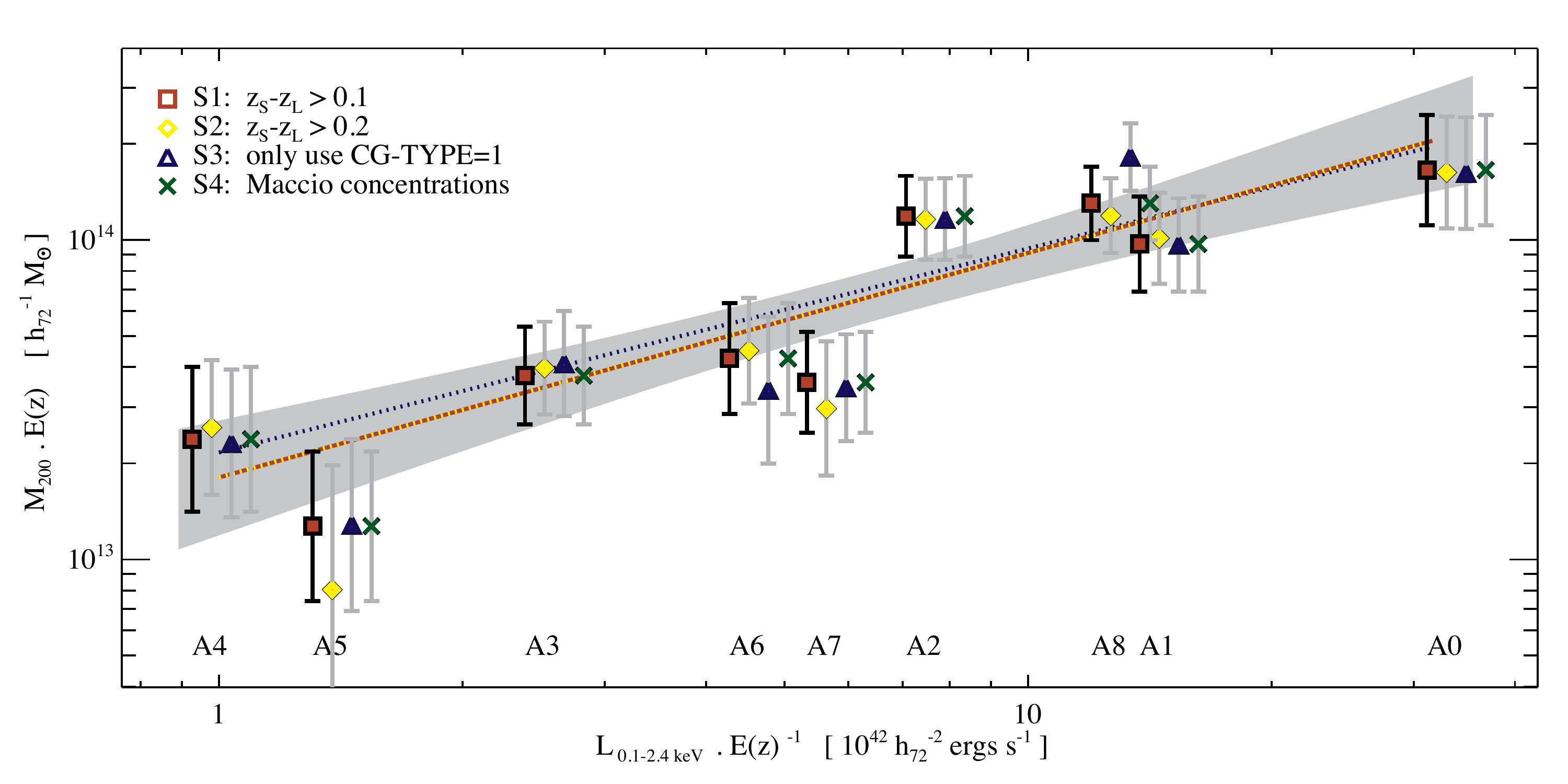}}
\caption{Tests for systematic errors. We implement four tests that are
  designed to check for errors in the foreground-background separation
  scheme, to probe the effects of mis-centering, and to quantify the
  impacts of variations in the mass-concentration relation. The
  lensing signal is re-computed for each test and the results are
  compared to the masses used to calculate
  $\mathcal{R}1_{L_{X}-M}$. The data points have been offset on the X
  axis by a constant value for visualization purposes. The grey shaded
  region shows the upper and lower envelope of the ensemble of lines
  with a slope and intercept that lie within the 68 percent confidence
  region of $\mathcal{R}1_{L_{X}-M}$. All the effects that are
  mentioned above are found to have a negligible impact on the
  results.}
\label{systematics}
\end{figure*}

We test each of these two schemes and the results are shown in Figure
\ref{systematics}. Choosing a value of $\delta z=0.1$ or $\delta
z=0.2$ has a negligible impact on $\rm{M}_{\rm 200}$.

\begin{figure*}[htb]
\centerline{\includegraphics[scale=0.56]{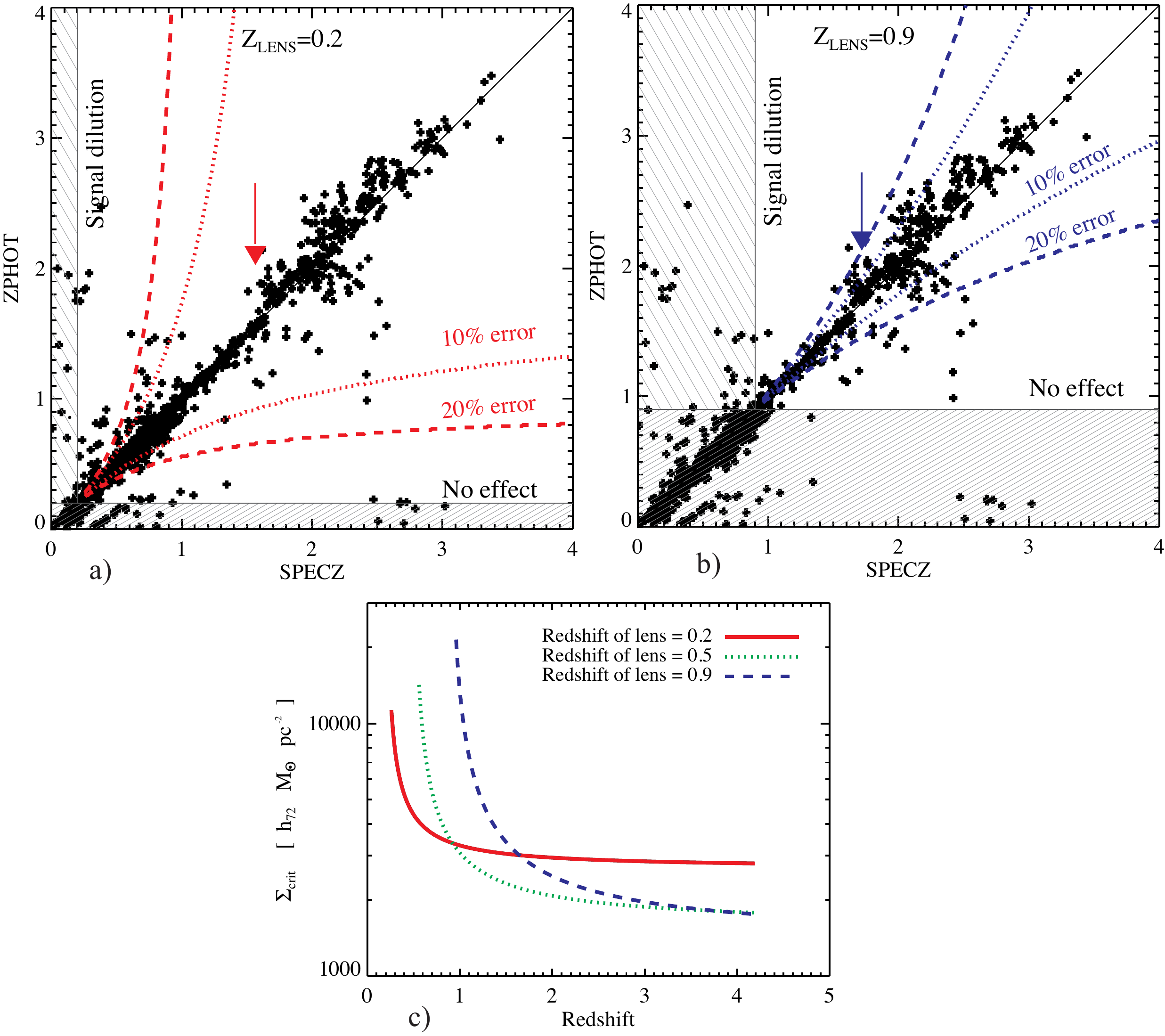}}
\caption{The effects of photometric redshifts errors on group-galaxy
  lensing signals. Panels a) and b) illustrate the quality of the
  photometric redshifts used in this paper by comparing them to a
  combined sample of 12367 spectroscopic redshifts from the zCOSMOS
  ``bright'' and ``faint'' programs. There are three ways in which
  photometric redshift errors can impact group-galaxy lensing
  signals. Firstly, any type of photometric error such that $z_{\rm
    phot}<z_{\rm lens}$ will have no effect on the signal because such
  objects are not included in the background selection (bottom hashed
  region). Secondly, photometric errors such that $z_{\rm phot}>z_{\rm
    lens}$ and $z_{\rm spec}<z_{\rm lens}$ will lead to a signal
  dilution (left hashed region). Finally, photometric errors such that
  $z_{\rm phot}>z_{\rm lens}$ and $z_{\rm spec}>z_{\rm lens}$ but
  $z_{\rm phot}\neq z_{\rm spec}$ will lead to a bias in
  $\Delta\Sigma$ because $\Sigma_{\rm crit}$ will be mis-estimated
  when transforming $\gamma$ into $\Delta\Sigma$. The dotted and
  dashed lines in panels a) and b) indicate where photoz errors lead to
  a $10\%$ and a $20\%$ error on $\Delta\Sigma$. As can be seen,
  group-galaxy lensing signals are increasingly insensitive to
  photometric redshift errors at higher source redshifts. This can be
  understood by looking at the variation of $\Sigma_{\rm crit}$ as a
  function of source redshift as shown in panel c). Indeed,
  $\Sigma_{\rm crit}$ varies strongly near $z_{\rm lens}$ but flattens
  considerably at $z_{\rm source}>2 z_{\rm lens}$. The arrows in
  panels a) and b) show where the mean COSMOS source redshift lies for
  lenses at $z_{\rm lens}$=0.2 and $z_{\rm lens}$=0.9. The bias in
  $\Delta\Sigma$ for group-galaxy lensing in COSMOS is estimated to be
  less than $7\%$ for lenses below $z=1$.}
\label{redshift_systematics}
\end{figure*}

In order to quantify the errors associated with (b) and (d), we make
use of the ensemble of spectroscopic redshifts that are currently
available from the zCOSMOS program. As illustrated in Figure
\ref{redshift_systematics}, source galaxies with $z_{\rm
  phot} > z_{\rm lens}$ and $z_{\rm spec} < z_{\rm lens}$ dilute the signal whereas galaxies with $z_{\rm
  phot} > z_{\rm lens}$ and $z_{\rm spec} > z_{\rm lens}$ but $z_{\rm
  phot}\neq z_{\rm spec}$ will introduce a bias in $\Delta\Sigma$
because $\Sigma_{\rm crit}$ will be mis-estimated when transforming
$\gamma$ into $\Delta\Sigma$. The correction factor for biases in the
photometric redshifts is noted $f_{bias}$ while $f_{boost}$ represents
a number greater than 1 that boosts the measured signal to compensate
for signal dilution. The true signal is related to the measured one
according to:

\begin{equation}
 \Delta\Sigma_{true} = \Delta\Sigma_{meas}\times f_{bias} \times f_{boost}. 
\label{eq:z_error}
\end{equation}

Using the ensemble of zCOSMOS spectra that are currently available, we
have estimated $f_{bias}$ and $f_{boost}$ for each of our nine
bins. The maximum bias that we find is a 7\% upwards correction on
$\Delta\Sigma$ for the last redshift bin which corresponds to a
revised mass of 8.5$\times$10$^{13}$ $h_{72}^{-1}$ M$_{\odot}$. This
is a small correction compared to our measured value of
7.6$^{+2.3}_{-1.8} \times10^{13}$ $h_{72}^{-1}$ M$_{\odot}$. In
conclusion, to the extent that we can estimate $f_{bias}$ and
$f_{boost}$ using the current zCOSMOS data, we find that the errors
due to imperfect photometric redshifts are below the statistical
uncertainties. It is important to note, however, that the zCOSMOS
spectroscopic sample may not be fully representative of the background
source population because of incompleteness. Thus it is possible that
our estimates of $f_{bias}$ and $f_{boost}$ are erroneous. For this
reason, we do not apply these correction factors to the data and
have simply listed the values of $f_{bias}$ and $f_{boost}$ in Table 1
as an indication of the probable systematic uncertainty due to
photometric redshift errors.

\subsection{The effects of mis-centering}

To test for mis-centering effects, we recalculate the lensing signals
using only systems with {\sc cg-type}=1 instead of using both {\sc
  cg-type}=1 and {\sc cg-type}=2. The results are shown in Figure
\ref{systematics} and we find that restricting our analysis to the
sub-sample of groups that have visually obvious central galaxies has
no impact on our estimates of $\rm{M}_{\rm 200}$.

\subsection{The mass concentration relation}

Testing for the effects of theoretical uncertainties in the
$M_{200}-C_{200}$ relation is beyond the scope of this paper, however,
one test we can perform is whether or not different $M_{200}-C_{200}$
relations affect our mass estimates. For this purpose, we compare two
recently derived $M_{200}-C_{200}$ relations, one from
\citet{Maccio:2007} and the second from \citet{Zhao:2008}. We compute
our lensing signals with each of these relations and show that the
results are largely unaffected by this test. Note that the agreement
in the $M_{200}-C_{200}$ relation from various authors is fairly good
in the mass and redshift regime of our group sample (the typical
fractional difference is 10\% to 20\%). However, at higher masses the
disagreement is larger and hence the assumed concentrations in H07 and
BA07 could represent a systematic error in the joint fit between
COSMOS and the cluster data.

\subsection{Conclusions regarding systematic uncertainties}

As demonstrated in Figure \ref{systematics}, all the effects that we
have tested for are largely negligible compared to the statistical
error. However, one aspect that we have not explored in this work is
the fact that each of our stacks contains a relatively small number of
groups. Indeed, the assumption of spherical symmetry will begin to
break down for stacks that only contain a small number of objects and
the weak lensing signal may be contaminated from projection effects
that have not averaged away. We have tried to limit this effect by
discarding all systems with visible projections along the
line-of-sight. In total, the various quality cuts described in
$\S$\ref{group_selection} are such that about 30\% of the initial
sample is rejected from the lens catalog. Most of these cuts were
linked to the quality of the X-ray data and to projection
effects. Additional {\sl XMM-Newton} and {\sl Chandra} data would
increase the size of our lens sample and would help reduce the
statistical error on the mass measurement.


\section{Summary and Discussion}\label{discussion}

In this work, we have used a sample of \ngroup X-ray detected galaxy
groups to investigate the scaling relation between total mass and
X-ray luminosity where ${\rm M}_{\rm 200}$ is derived via weak
gravitational lensing. In the following paragraphs, we present a
summary and discussion of our main results.  \vspace{0.05 in}

{\bf The combination of group and cluster data}. The COSMOS group
catalog spans an approximate mass and redshift range $M_{200} \sim
10^{13} \mass$ to $M_{200} \sim 10^{14} \mass$ with $0.2 < z < 0.9$, a
new parameter space in terms of weak lensing-based mass measurements
of X-ray detected groups and clusters of galaxies. When appropriately
scaled for self-similar redshift evolution, the COSMOS data, alongside
previously published results display a remarkable power-law relation
that spans over three orders of magnitude in $L_X E(z)^{-1}$ and two
orders of magnitude in $M_{200}E(z)$. The COSMOS data alone are well
fit by a power law, $M_{200} \propto (L_{X})^{\alpha}$, with a slope
of \cosalpha. By combining with previously published cluster data, we
derive a tighter constraint on the slope, \henkalpha. This is
inconsistent at the \sssig\ level with the self-similar prediction of
$\alpha=0.75$.  Note that the combination of group and cluster data
greatly helps constrain the slope of the \mlx relation but the
determination of the normalization is at present limited by the
accuracy of weak lensing measurements. Improvements in weak lensing
methods and larger group and cluster samples will be necessary in
order to improve constraints on the normalization of the \mlx
relation.\vspace{0.05 in}

{\bf Comparison with previous lensing results}. Our analysis compares
best with the local SDSS results from R08 who find $L_{\rm X} \propto
(M_{200})^{\beta}$ with $\beta=1.65 \pm 0.13$. The inverse of their
slope, $1/\beta = 0.61 \pm 0.048$ is in excellent agreement with both
$\mathcal{R}1_{M-L_{X}}$ and $\mathcal{R}2_{M-L_{X}}$. As described in
their paper, because the R08 results are binned by richness, the slope
of their relation is sensitive to the correlation coefficient between
richness and \lx at fixed mass, $r_{N,L|M}$ (see $\S$4 in R08). A
value of $r_{N,L|M}=\pm 0.7$ would change their slope to $0.68 \pm
0.061$ and $0.54\pm 0.038$ respectively. These slopes would still be
in relatively good agreement with the COSMOS results but much larger
values than $r_{N,L|M}=\pm 0.7$ can be ruled out. Small values for
$r_{N,L|M}$ are also favored by the analysis of \citet[][]{Rozo:2008}
who find a value of $r_{N,L|M} \sim 0.05$ (Rozo priv. comm.).

\begin{figure*}[htb]
\centerline{\includegraphics[angle=90,scale=0.57]{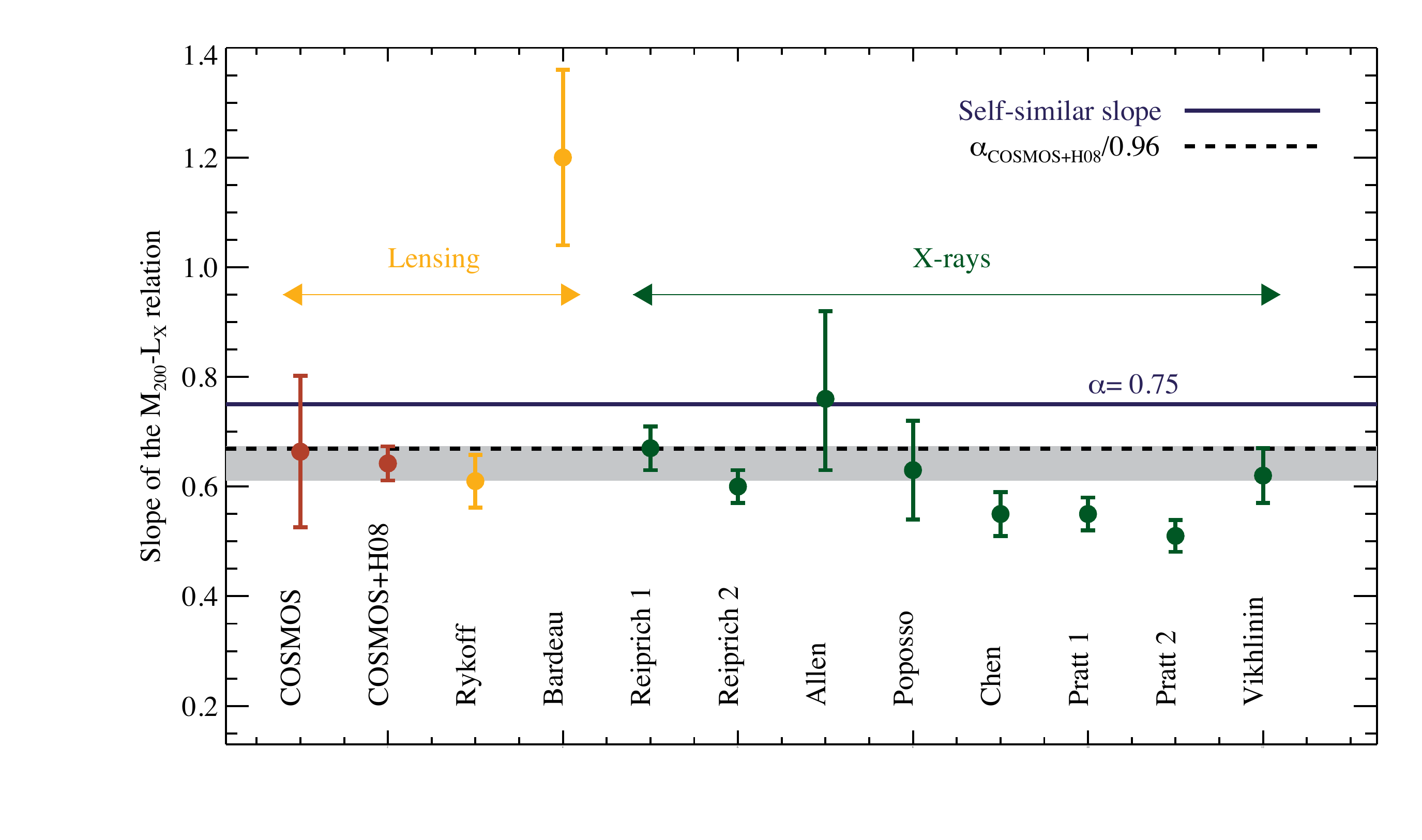}}
\caption{A comparison of the slope of the \mlx relation obtained by
  various authors. From left to right we show $\mathcal{R}1_{M-L_{X}}$
  and $\mathcal{R}2_{M-L_{X}}$ (this work) followed by the lensing
  based results of \protect\citet{Rykoff:2008} and
  \protect\citet{Bardeau:2007} and then by X-ray-based results
  \protect\citep[][]{Reiprich:2002,Allen:2003,Popesso:2005,Chen:2007,Pratt:2008,Vikhlinin:2009}. The
  solid blue shows the self-similar prediction for the slope which is
  $\alpha=0.75$. The grey shaded region indicates the one sigma errors
  for $\mathcal{R}2_{M-L_{X}}$. With the exception of the BA07
  results, the lensing and the X-ray results are in good agreement
  with an average slope of $\alpha \sim 0.64$. Note that because of
  scatter in the \lxm relation and a halo mass function with a varying
  slope, the lensing and the X-ray results are not directly
  comparable. The dashed black line indicates the value of the
  COSMOS+H08 data point corrected for the difference between $1/\beta$
  and $\alpha_{lensing}$ assuming a scatter of $\sigma_{\ln M}=0.25$
  (see derivations in Appendix). \hspace{2cm}}
\label{comp_slope}
\end{figure*}

BA07 have also published a $M-L_{\rm X}$ relation but find a slope of
$1.20 \pm 0.16$ that is inconsistent with the COSMOS value at the
3.4$\sigma$ level. We suspect that this disagreement stems from an
underestimate of the BA07 masses, in particular at the low-mass end of
the cluster sample. Indeed, H07 have analyzed almost exactly the same
set of clusters but find a slope and normalization that is in better
agreement with both this work and R08 (see Figure \ref{lx_m} and also
Figure \ref{comp_slope}). Nevertheless, despite systematic differences
between the mass estimates of H07 and BA07, the combined relations
using either data set with COSMOS are almost identical. Manifestly,
the addition of group data has significantly reduced the impact of
cluster mass uncertainties on the \mlx relation. \vspace{0.05 in}

{\bf Comparison with previous X-ray results}. The comparison of the
normalization of the \mlx relation derived with lensing on the one
hand and with X-rays on the other hand is of great interest because it
has the potential to reveal systematic biases in X-ray-based cluster
mass estimates (\eg\ due to non-thermal processes such as turbulence
and cosmic-rays). However, the comparison of the normalization between
the \mlx relation and the \lxm relation is complex because it depends
on $\sigma_{\ln M}$ and the slope of the halo mass function (see
Appendix) and we leave this aspect for future work. Therefore, we
mainly focus on a comparison of the slopes. For this purpose, we have
compiled a list of the slopes of the \lxm relation as determined by
various X-ray studies. When authors have stated their results in terms
of the \lxm relation, we have inverted the slope in order to compare
with our results\footnote[8]{Errors on $y=\frac{1}{x}$ have been
  computed as $\Delta y = \left( \frac{1}{x} \right)^2 \times \Delta
  x$}. In the Appendix we show that although the slope of $P(L_{X}|M)$
is not exactly equal to the inverse of the slope of $P(M|L_{X})$, the
difference is small (also see Figure \ref{bias_illustration}).

A representative (but not exhaustive) list of X-ray-based results is
the following:

\begin{itemize}
\item \citet[][]{Reiprich:2002}: $L_{\rm X} \propto M_{200} ^{1.496\pm
    0.089}$ or $L_{\rm X} \propto M_{200} ^{1.652\pm 0.085}$ depending
  on the fitting method (extended sample, Table 7 of their paper). The
  inverse slopes are $0.67\pm 0.04$ and $0.60\pm 0.03$, respectively.

\item \citet[][]{Allen:2003}: $M_{200} \propto L_{\rm
    X}^{0.76_{-0.13}^{+0.16}}$.
\item \citet[][]{Popesso:2005}: $L_{\rm X} \propto M_{200} ^{1.58\pm
    0.23}$. The inverse slope is $0.63\pm 0.09$.
\item \citet[][]{Chen:2007} : $L_{\rm X} \propto M_{200} ^{1.82\pm
    0.13}$. The inverse slope is $0.55\pm 0.04$.
\item \citet[][]{Pratt:2008}: $L_{1} \propto M_{Y}^{1.81 \pm 0.10}$ or
  $L_{1} \propto M_{Y}^{1.96 \pm 0.11}$ depending on the fitting
  method (Table 2 in their paper). The inverse slopes are
  $0.55 \pm 0.03$ and $0.51 \pm 0.029$.
\item \citet[][]{Vikhlinin:2009}: $L_{\rm X} \propto M_{200} ^{1.61\pm
    0.14}$. The inverse slope is $0.62\pm 0.05$.
\end{itemize}

Figure \ref{comp_slope} shows a comparison between X-ray based
estimates of the slope of the \mlx relation and lensing-based
results. As can be seen in this Figure, most of the lensing and the
X-ray results are in excellent agreement with an average slope of
$\alpha \sim 0.64$.

\vspace{0.05 in}

{\bf Evolution of scaling relations}. Large surveys that will probe
clusters up to $z=1$ will need a precise understanding of the redshift
evolution in mass-observable relations. In this paper, we have shown
that weak gravitational lensing is capable of meeting this challenge
and we have tested several evolution scenarios from $z\sim 0.2$ to
$z\sim 0.9$. Our results are consistent with the evolution rate
predicted by self-similarity but our errors are still relatively large
due to our small sample size at high redshift. Additional X-ray data
would help improve the precision of this measurement. More precise
measurements of the redshift evolution of X-ray scaling laws would
also help constrain the physical processes that govern the heating and
cooling of the ICM. \vspace{0.05 in}

{\bf Self-calibration methods}. Sufficiently large cluster surveys may
be able to deal with the evolution and scatter in mass-observable
relations by internally calibrating for these uncertainties \citep[the
so called ``self-calibration
method'',][]{Levine:2002,Hu:2003a,Majumdar:2004,Wang:2004b,Lima:2005}. This
method treats all uncertainties as free parameters to be fitted along
with the desired cosmological parameters. However, there are several
drawbacks to this method. The first is that treating systematic errors
in this manner weakens the final statistical constraints. The second
is that self-calibration requires a parametric form for the scatter
and evolution of scaling relations. Bias is introduced if this
parametrization is incorrect.

This paper has showed that weak gravitational lensing can help
constrain the actual form of the mass-observable relations as well as
their evolution, even out to high redshifts ($z<1$). Having a (direct
observational) external constraint will help boost the accuracy
achievable with self-calibration methods by reducing the number of
parameters and by pinning down the correct parametric form. One
important ingredient in the \mlx relation, and for self-calibration
exercises, is the scatter $\sigma_{\ln M}$. Stacked weak lensing
measurements are not suitable for measuring the scatter and so
individual lensing detections will be necessary for this
task. Consequently, estimates of the scatter with lensing will be
limited to high masses and moderate redshifts (see Figure
\ref{james_plot}). Space-based data will be optimal for scatter
studies because high source densities will increase the mass and
redshift range for which clusters can be directly probed. The
challenge for lensing-based estimates of the scatter will be to ensure
that the lensing errors are smaller than the intrinsic scatter, a
condition that is probably not achieved at present.

Nevertheless, one interesting point to note is that scatter will
introduce a small amount of curvature in stacked lensing based
measurements of the \mlx relation (see Figure
\ref{bias_illustration}). Therefore, it is possible that the scatter
could be measured with future data by constraining the amount of
curvature in \mlx at high halo masses.\vspace{0.05 in}

In conclusion, the field of weak gravitational has started to become a
truly competitive tool for calibrating the relation between the total
mass of groups and clusters of galaxies and their baryonic tracers,
over a wide range of masses ($\rm{M}_{\rm 200} \sim
10^{13.5}$--$10^{15.5} \mass$) and up to $z=1$. At present, the slope
of the \mlx relation is constrained by lensing with a statistical
significance that is comparable to X-ray studies (at the $\sim 5\%$
level). Although further work will be necessary in order to compare
both the slope and the normalization, it is encouraging to note that
both lensing and X-rays studies are already in good agreement with
respect to the slope of the \mlx relation ($\alpha \sim 0.64$). The
observational foundation for the calibration of the mass-observable
relations that are essential for cosmological studies with clusters of
galaxies is clearly growing firmer.\vspace{0.05 in}


\acknowledgments
\noindent {\bf Acknowledgments}

We thank Uros Seljak, Reiko Nakajima, Henk Hoekstra, Rachel
Mandelbaum, Chris Hirata, Masahiro Takada, and Satoshi Miyazaki for
useful discussions. We thank Florian Pacaud for providing X-ray data
for this analysis. We are also grateful to Eduardo Rozo and to Joel
Berg\'{e} for providing comments on the manuscript and to an anonymous
referee for insightful comments. AL acknowledges support from the
Chamberlain Fellowship at LBNL and from the Berkeley Center for
Cosmological Physics. JPK acknowledges CNRS and CNES for support. JPK
and HJMCC acknowledge support from the research grant ANR-07-
BLAN-0228. The HST COSMOS Treasury program was supported through NASA
grant HST-GO-09822. We wish to thank Tony Roman, Denise Taylor, and
David Soderblom for their assistance in planning and scheduling of the
extensive COSMOS observations.  We gratefully acknowledge the
contributions of the entire COSMOS collaboration consisting of more
than 70 scientists.  More information on the COSMOS survey is
available at {\bf \url{http://cosmos.astro.caltech.edu/}}. It is a
pleasure the acknowledge the excellent services provided by the NASA
IPAC/IRSA staff (Anastasia Laity, Anastasia Alexov, Bruce Berriman and
John Good) in providing online archive and server capabilities for the
COSMOS datasets.


\appendix

In this Appendix, we derive the relation between
$P(\mathrm{M}|\mathrm{L}_X,z)$ and $P(\mathrm{L}_X|M,z)$ and we show
that a correction needs to be applied when comparing the slopes of the
mean log luminosity and the mean log mass over a wide range in halo
mass. The slope of the mean relation associated with
$P(\mathrm{M}|\mathrm{L}_X,z)$ is noted $\alpha_{lensing}$ and the
slope of $P(\mathrm{L}_X|M,z)$ is noted $\beta$. Partial aspects of
this derivation can also be found in the Appendix of
\citet[][]{Mandelbaum:2007}.

Let $m \equiv \ln \mathrm{M}$ and $l\equiv \ln \mathrm{L}_{X}$ where M
represents halo mass. Following \citet[][]{Stanek:2006}, we assume
that the conditional probability distribution of the luminosity given
the mass is log-normal. In this case, $P(l|m,z)$ is Gaussian and we
can write that:

\begin{equation}\label{ap_eqa1}
  P(l|m,z) = \frac{1}{\sigma_{l}\sqrt{2\pi}} \times \mathrm{exp} \left( \frac{-[l-l_0(m,z)]^2}{2\sigma_l^2} \right),
\end{equation}

\noindent where $l_0(m,z)$ is the mean log luminosity and $\sigma_l$
is the scatter (also noted $\sigma_{\ln L_{X}}$). We assume that the
scatter varies neither with mass, nor redshift ($\sigma_{l}$ is
constant) and that the mean log luminosity follows a power-law scaling
relation with mass and with self-similar redshift evolution:

\begin{equation}\label{ap_eqa2}
  \langle \ln L_{\rm X}\rangle  \equiv l_{0}(m,z) = \beta m + (1+\beta)\ln E(z) + B.
\end{equation}

Let $n(M)$ represent the number of dark matter halos per unit volume
with mass less than M. Locally, the differential mass function is a
power-law of the form ${\rm d}n/{\rm d}M \propto
M^{-\gamma}=e^{-\gamma m}$. The probability of observing a halo of
log-mass $m$ is $P(m)={\rm d}n/{\rm d}\ln M \propto e^{-(\gamma-1) m}$.

The weak lensing signal of halos stacked according to \lx depends on
$P(\mathrm{M}|\mathrm{L}_X,z)$, the conditional probability
distribution of the mass given the luminosity. Using Bayes theorem and
ignoring those terms that only contribute to the overall normalization
of $P(m|l,z)$, we can write that

\begin{equation}\label{ap_eqa3}
  P(m|l,z) \propto P(l|m,z) P(m,z)
  \propto \mathrm{exp} \left( \frac{-[l-l_0(m,z)]^2}{2\sigma_l^2} -(\gamma-1) m \right).
\end{equation}

By using Equation \ref{ap_eqa2} to develop the expression within the
exponential and by completing the square, we obtain that

\begin{equation}
  P(m|l,z) \propto \mathrm{exp} \left( \frac{-[m-m_0(l,z)]^2}{2(\sigma_l \beta^{-1})^2} \right),
\end{equation}

\noindent where $ \langle \ln M\rangle \equiv m_{0}(l,z)=(1/\beta)
l-(1+1/\beta)\ln E(z)- B/\beta -\sigma_{m}^2(\gamma-1)$. In other
terms, $P(m|l,z)$ is Gaussian and $P(\mathrm{M}|\mathrm{L}_X,z)$ is
log-normal with a dispersion in mass equal to
$\sigma_{m}=\sigma_{l}/\beta$ and with a mean log mass that follows a
power-law of with a slope of $\alpha_{lensing}=1/\beta$. The mean
log-mass, $m_{0}(l,z)$, can be obtained by solving for the mass in
Equation \ref{ap_eqa2} with the addition of an extra term. In this
sense, there is a bias between $m_{0}(l,z)$ and the true mean mass
which is equal to $b_{m}=\sigma_{m}^2(\gamma-1)$ and this bias scales
linearly with $\gamma-1$ where $\gamma$ is the slope of the mass
function. Because, $(\gamma-1)$ is always positive, this bias causes
$m_{0}(l,z)$ to be biased low relative to the true mean mass.

Both mass and luminosity are often expressed in units of logarithm to
the base 10 rather than in units of natural logarithm. In this case,
the scatter in mass at fixed luminosity is equal to $\sigma_{\log_{10}
  M} \equiv \sigma_{m10}=\sigma_{m}/\ln(10)$ and the bias is equal to
$b_{m10}=\sigma_{m10}^2(\gamma-1)
\ln(10)=\sigma_{m}^2(\gamma-1)/\ln(10)$. Note that there is an extra
factor of $\ln(10)$ in the expression of the bias due to the fact that
$P(\log_{10} M)={\rm d}n/{\rm d}\log_{10} M \propto M^{-(\gamma-1)}=e^{-(\gamma-1)
  \ln(10) \log_{10}(M)}$.

We have demonstrated that $\alpha_{lensing}=1/\beta$ when the halo mass function
is locally a power-law. In reality, the slope of the mass function
varies with both mass and redshift. At $z=0.2$, the slope of the mass
function varies from $\gamma \sim 2$ at $M_{200} \sim 10^{13}~\mass$
to $\gamma \sim 5$ at $M_{200} \sim 10^{15}~\mass$. The fact that the
bias depends on $\gamma$ will result in a {\em change in slope} and
$\alpha_{lensing}=1/\beta$ will no longer be valid. Instead, there will be a
correction factor between the slope derive by stacked weak lensing,
$\alpha_{lensing}$, and $1/\beta$ and we will now show how this
correction factor can be roughly estimated.

Let $m_{1}$ and $m_{2}$ represent two distinct halo masses in units of
logarithm to the base 10. The local slope of the mass function at each
of these masses is noted $\gamma_1$ and $\gamma_2$ and the biases are
noted $b_{m1}=\sigma_{m}^2\gamma_1/\ln(10)$ and
$b_{m2}=\sigma_{m10}^2\gamma_2/\ln(10)$. The slope that is derived via
stacked weak lensing over the mass range $[m_{1},m_{2}]$ is equal to
$\alpha_{lensing}=F/\beta$ where $F=1-(b_{m1}-b_{m2})/(m_1-m_2)$. For
$z=0.2$, $\sigma_{m}=0.25$ ($\sigma_{m10} \sim 0.109$), $M_{1} \sim
10^{13}~\mass$ and $M_{2} \sim 10^{15}~\mass$, we have $b_{m1} \sim
0.03$ and $b_{m2} \sim 0.11$ (when $m$ is expressed in log$_{10}$
units). The difference between $1/\beta$ and $\alpha_{lensing}$ over
this mass range is therefore roughly $\alpha_{lensing} \sim
0.96/\beta$.



\bibliographystyle{apj}


\end{document}